\documentclass[journal]{IEEEtran}
\usepackage{cite}
\usepackage{amsmath,amssymb,amsfonts}
\usepackage{graphicx}
\usepackage{textcomp}
\usepackage{epstopdf}
\usepackage[caption=false, font=footnotesize]{subfig}
\usepackage{times}
\usepackage{algorithm}
\usepackage{algorithmic}
\usepackage{multirow}
\usepackage[table,xcdraw]{xcolor}
\usepackage{xcolor}
\usepackage[draft]{hyperref}
\ifCLASSINFOpdf
\else
\fi
\begin{document}
%
\title{Affective EEG-Based Person Identification Using the Deep Learning Approach}
\author{Theerawit Wilaiprasitporn, Apiwat Ditthapron, Karis Matchaparn, Tanaboon Tongbuasirilai, \\Nannapas Banluesombatkul and Ekapol Chuangsuwanich
\thanks{This work was supported by The Thailand Research Fund under Grant MRG6180028.}
\thanks{T. Wilaiprasitporn and N. Banluesombatkul are with Bio-inspired Robotics and Neural Engineering Lab, School of Information Science and Technology, Vidyasirimedhi Institute of Science \& Engineering, Rayong, Thailand (e-mail: theerawit.w@vistec.ac.th).}
\thanks{A. Ditthapron is with the Computer Department, Worcester Polytechnic Institute, Worcester, MA, USA.}
\thanks{K. Matchaparn is with the Computer Engineering Department, King Mongkut's University of Technology Thonburi, Bangkok, Thailand.}
\thanks{T. Tongbuasirilai is with Department of Science and Technology, Link\"{o}ping University, Sweden}
\thanks{E. Chuangsuwanich is with the Computer Engineering Department, Chulalongkorn University, Bangkok, Thailand.}
}

%
%

\markboth{Journal of \LaTeX\ Class Files,~Vol.~14, No.~8, August~2015}%
{Shell \MakeLowercase{\textit{et al.}}: Bare Demo of IEEEtran.cls for IEEE Journals}
%



\maketitle

\begin{abstract}
Electroencephalography (EEG) is another mode for performing Person Identification (PI). Due to the nature of the EEG signals, EEG-based PI is typically done while the person is performing some kind of mental task, such as motor control. However, few works have considered EEG-based PI while the person is in different mental states (affective EEG). The aim of this paper is to improve the performance of affective EEG-based PI using a deep learning approach. \textcolor{red}{We proposed a cascade of deep learning using a combination of Convolutional Neural Networks (CNNs) and Recurrent Neural Networks (RNNs)}. CNNs are used to handle the spatial information from the EEG while RNNs extract the temporal information. \textcolor{red}{We evaluated two types of RNNs, namely, Long Short-Term Memory (CNN-LSTM) and Gated Recurrent Unit (CNN-GRU).
} The proposed method is evaluated on the state-of-the-art affective dataset DEAP. The results indicate that CNN-GRU and CNN-LSTM can perform PI from different affective states and reach up to 99.90--100\% mean Correct Recognition Rate (CRR), significantly outperforming a support vector machine (SVM) baseline system that uses power spectral density (PSD) features.  Notably, the 100\% mean \emph{CRR} comes from only 40 subjects in DEAP dataset. To reduce the number of EEG electrodes from thirty-two to five for more practical applications, the frontal region gives the best results reaching up to 99.17\% CRR (from CNN-GRU). Amongst the two deep learning models, we find CNN-GRU to slightly outperform CNN-LSTM, while having faster training time. \textcolor{red}{Furthermore, CNN-GRU overcomes the influence of affective states in EEG-Based PI reported in the previous works.}
\end{abstract}

\begin{IEEEkeywords}
Electroencephalography, Personal identification, Biometrics, Deep learning, Affective computing, Convolutional neural networks, Long short-term memory, Recurrent neural networks
\end{IEEEkeywords}

\section{Introduction}
\label{sec:introduction}
\IEEEPARstart{I}{n} today\textquotesingle s world of large and complex data-driven applications, research engineers are inspired to incorporate multiple layers of artificial neural networks or deep learning (DL) techniques into health informatic-related studies such as bioinformatics, medical imaging, pervasive sensing, medical informatics and public health \cite{Rav`i2017}. Such studies also include those relating to frontier neural engineering research into brain activity using the non-invasive measurement technique called electroencephalography (EEG). The fundamental concept of EEG involves measuring electrical activity (variation of voltages) across the scalp. The EEG signal is one of the most complex in health data and can benefit from DL techniques in various applications such as insomnia diagnosis, seizure detection, sleep studies, emotion recognition, and Brain-Computer Interface (BCI) \cite{Movahedi2017,Shahin2017,Lu2017,Zhang2017,Tabar2016,Schirrmeister2017}. However, EEG-based Person Identification (PI) research using DL is scarcely found in literature. Thus, we are motivated to work in this direction.

EEG-based PI is a biometric PI system--fingerprints, iris, and face for example. EEG signals are determined by a person's unique pattern and influenced by mood, stress, and mental state \cite{qgui2014}. EEG-based PI has the potential to protect encrypted data under threat. Unlike other biometrics, EEG signals are difficult to collect surreptitiously, since they are concealed within the brain \cite{Mal2016}. Besides the person's unique pattern, a passcode or pin can also be recognized from the same signal, while having a low chance of being eavesdropped. EEG signals also leave no heat signal or fingerprint behind after use.

The PI process shares certain similarities with the person verification process, but their purposes are different. Person verification validates the biometrics to confirm a person's identity (one-to-one matching), while PI uses biometrics to search for an identity match (one-to-many matching) on the database \cite{vacca2007biometric}. EEG-based PI system development has dramatically increased in recent years \cite{Campisi2014,Yang2017}. Motor tasks (eye closing \cite{Ma}, hands movement \cite{patel}, etc.), visual stimulation \cite{Min2017,Chen2016,Das} and multiple mental tasks such as mathematical calculation, writing text, and imagining movements (\cite{Das2}) are three major tasks in stimulating brain responses for EEG-based PI \cite{DelPozo-Banos2014}. To identify a person, it is very important to investigate the stimulating tasks which can induce personal brain response patterns. Moods, feelings, and attitudes are usually related to personal mental states which react to the environment. \textcolor{red}{However, emotion-elicited EEG has been rarely investigated to perform person identification. There are several reports on affective EEG-based PI; one with a small affective dataset \cite{DelPozo-Banos2015}, one reaching less than a 90\% mean Correct Recognition Rate \emph{(CRR)}\cite{li2017} and another using the same dataset as our works with PSD-based feaures reached up to 97.97\% mean CRR.} Thus, the aim of this paper is to evaluate the usability of EEG from elicited emotions for person identification applications. The study of affective EEG-based PI can help us gain a greater understanding concerning the performance of personal identification among different affective states. This study mainly focuses on the state-of-the-art EEG dataset for emotion analysis named DEAP \cite{Koelstra2012}.

\textcolor{red}{A recent critical survey on the usability of EEG-based PI resulted in several major signal processing techniques to help perform  feature extraction and classification \cite{Yang2017}.} Power Spectral Density (PSD) methods \cite{Poulos1999,Palaniappan2007,LaRocca2014,Safont2012}, the Autoregressive Model (AR) \cite{Poulos1999a,Paranjape2001,Riera2008,Campisi2011,Maiorana2016,Dan2013}, Wavelet Transform (WT) \cite{Gupta2009,yang2013} and Hilbert-Huang Transform (HHT) \cite{Kumari2014,Yang2014} are useful for feature extraction. For feature classification, k-Nearest Neighbour (k-NN) algorithms \cite{Yazdani2008,Su2010}, Linear Discriminant Analysis (LDA) \cite{Palaniappan2006,Lee2013}, Artificial Neural Networks (ANNs) with a single hidden layer \cite{Poulos1999,Palaniappan2004,Palaniappan2011,Gui2014} and kernel methods \cite{Jian-feng2010,Ashby2011} are popular techniques. \textcolor{red}{In this study, we propose a DL technique to perform both the feature extraction and classification tasks.} The proposed DL model is a cascade of the CNN and GRU. CNN and GRU are supposed to capture spatial and temporal information from EEG signals, respectively. \textcolor{red}{A similar cascade model based on CNN and LSTM has recently been applied in a motor imagery EEG classification aiming at BCI applications, however, they did not study GRUs to perform this task \cite{Zhang2017}.}

\textcolor{red}{The main contribution of this investigation can be summarized as follows: 
\begin{itemize}
\item We propose an effective EEG-based PI using a DL approach, which has not been investigated previously to any extent.
\item The proposed approach overcomes the influence of affective states in using EEG for PI task.
\item Using our proposed technique, we investigate whether any EEG frequency bands outperform others to perform affect EEG-based PI task. 
\item We performed extensive experimental studies using a set of five EEG electrodes from different scalp areas. Specifically, these studies reports the feasibility of the proposed technique to handle real world scenarios.
\item We provide the performance comparison of the proposed cascade model (CNN-GRU) against a spatiotemporal DL model (CNN-LSTM), and other systems proposed in literature \cite{li2017,Banos}.
\end{itemize}
In summary, the experimental results guarantee that the CNN-GRU converges faster than CNN-LSTM while having a slightly higher mean CRR, especially when using a small amount of electrode. Furthermore, CNN-GRU overcomes the influence of affective states in EEG-Based PI reported in the previous works \cite{Banos, Arnau}.
}

The structure of this paper is as follows. Sections II and III present the background and methodology, respectively. The results are reported in Section IV. Section V discusses the results from the experimental studies. Moreover, the beneficial points are highlighted for comparison over previous works for further investigation. Finally, the conclusion is presented in Section VI.

\section{The Deep Learning Approach To EEG}
There has been a surge of deep learning-related methods for classification of EEG signals in recent years. Since EEG signals are recordings of biopotentials across the scalp over time, researchers tend to use DL architectures for capturing both spatial and temporal information. A cascade of CNN, followed by an RNN, often an LSTM, is typically used. These cascade architectures work according to the nature of neural networks, where the proceeding layers function as feature extractors for the latter layers. 

\textcolor{red}{CNN are often used as the initial layers of deep learning architectures in order to extract meaningful patterns or features. The key element of CNN is the convolution operation using small filter patches (kernels). These filters are able to automatically learn local patterns which can be combined together to form more complex features when stacking multiple CNN layers together. Within the stack of convolution layers, pooling layers are often placed intermittently. The pooling layers subsample the output of the convolution layers by outputting only the maximum value for each small region. The subsampling allows the convolution layer after the pooling layer to work on a different scale than the layers before it. These features learned from the CNN can be used as input to other network structures to perform sophisticated tasks such as object detection or semantic segmentation \cite{objectdetection}.}

For EEG signals, it makes sense to feed the local structures learned by the CNN to LSTMs, which can better handle temporal information. Zhang et al. also tried a 3D CNN to exploit the spatiotemporal information directly within a single layer. However, the results were slightly behind a cascade of the CNN-LSTM model \cite{Zhang2017}. This might be due to the fact that LSTMs are often better at handling temporal information since they can choose to remember and discard information depending on the context.

\begin{figure}[t] 
    \centering
  \subfloat[GRU]{
       \includegraphics[width=0.48\linewidth]{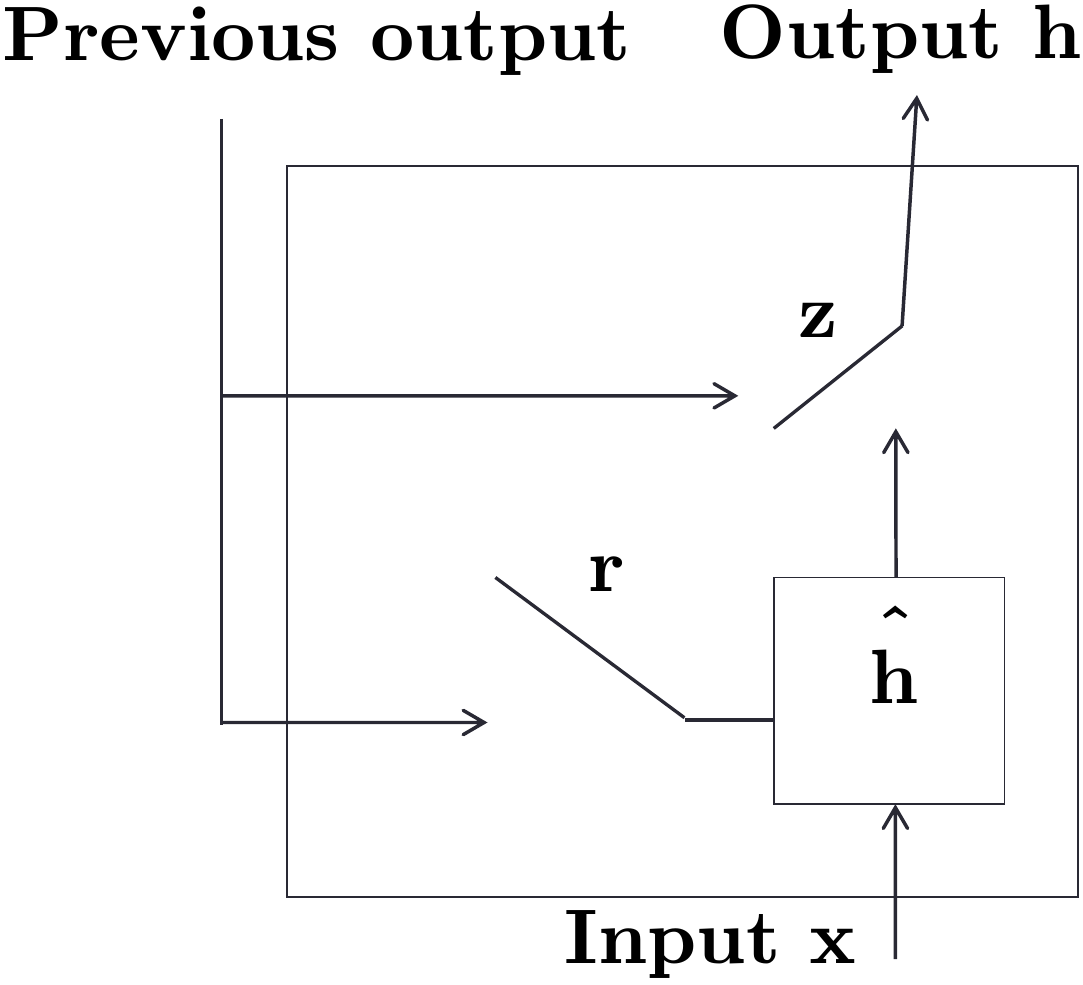}}
    \label{shows the interconnections of a GRU unit.}
  \subfloat[LSTM]{
        \includegraphics[width=0.48\linewidth]{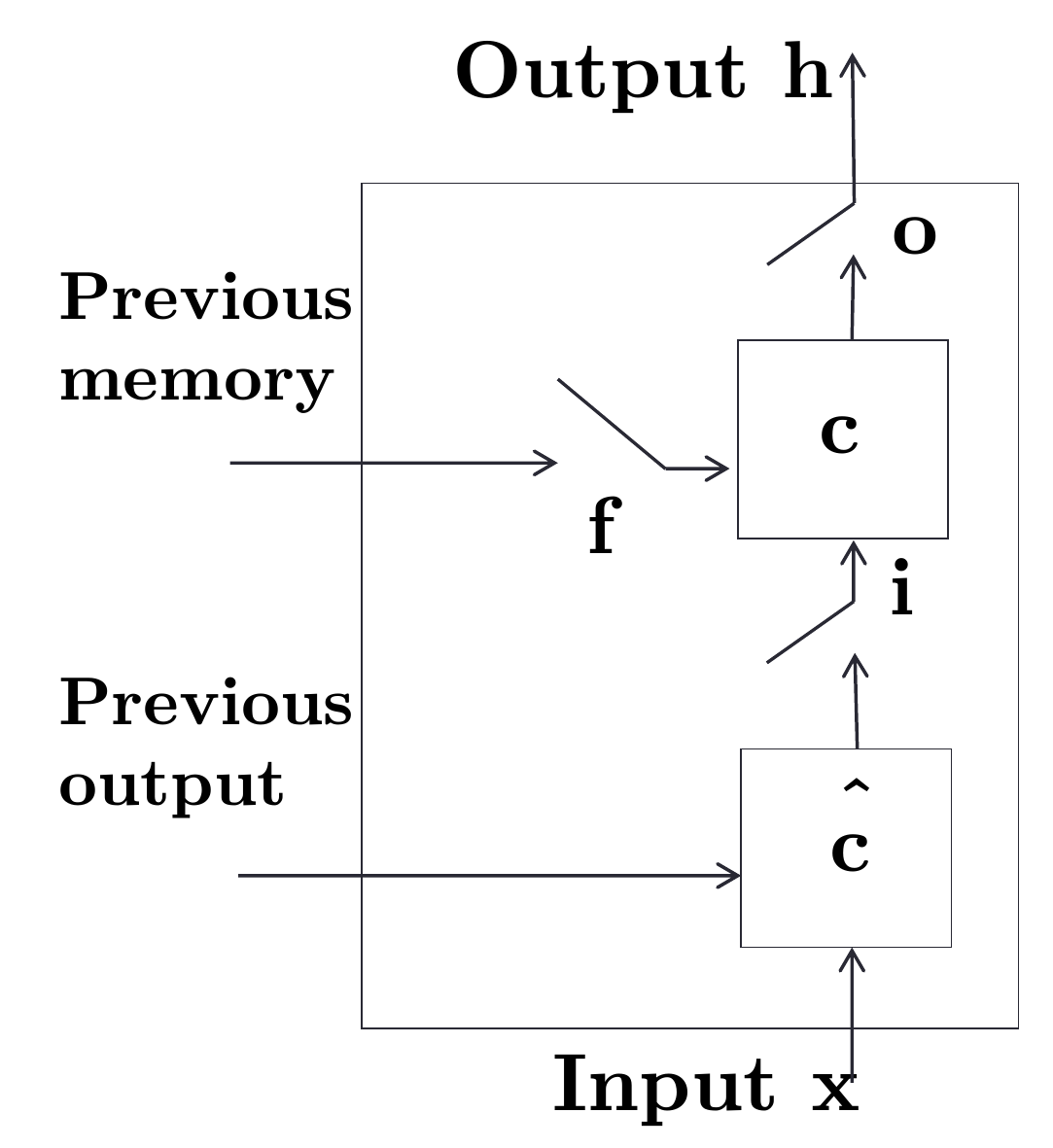}}
    \label{shows the interconnections of a LSTM unit.}
  \caption{Comparison between GRU and LSTM structures and their operations}
  \label{RNN}
\end{figure}

Another type of recurrent neural network called the Gated Recurrent Unit (GRU) has also been proposed as an alternative to the LSTM \cite{Cho2014}. The GRU can be considered as a simplified version of the LSTM. GRUs have two gates (reset and update) instead of three gates as in the LSTMs. GRUs directly output the captured memory, while LSTMs can choose not to output its content due to the output gate. \autoref{RNN} (a) shows the interconnections of a GRU unit. Just as a fully connected layer is composed of multiple neurons, a GRU layer is composed of multiple GRU units. Let $\mathbf{x}_t$ be the input at time step t to a GRU layer. The output of the GRU layer, $\mathbf{h}_t$, is a vector composing the output of each individual unit $h_t^j$, where $j$ is the index of the GRU cell. The output activation is a linear interpolation between the activation from the previous time step and a candidate activation, $\hat{h}^j_t$.

\begin{equation} \label{eqg1}
h_t^j = (1-z_t^j)h_{t-1}^j + z_t^j\hat{h}^j_t
\end{equation}
where an update gate, $z_t^j$, decides the interpolation weight. The update gate is computed by
\begin{equation} \label{eqg2}
z_t^j = F^j( W_z \mathbf{x}_t + U_z \mathbf{h}_{t-1})
\end{equation}
where $W_z$ and $U_z$ are trainable weight matrices for the update gate, and $F^j()$ takes the $j$-th index and pass it through a non-linear function (often a sigmoid).  The candidate activation is also controlled by an additional reset gate, $\mathbf{r}_t$, and computed as follows:
\begin{equation} \label{eqg3}
\hat{h}^j_t = G^j( W\mathbf{x}_t + U (\mathbf{r}_t \odot \mathbf{h}_{t-1}))
\end{equation}
where $\odot$ represents an element-wise multiplication, and $G^j()$ is often a tanh non-linearity. The reset gate is computed in a similar manner as the update gate:
\begin{equation} \label{eqg4}
r_t^j = F^j( W_r \mathbf{x}_t + U_r \mathbf{h}_{t-1})
\end{equation}

On the other hand, LSTMs have three gates, input, output, and forget gates which are denoted as $i_t^j$, $o_t^j$, $f_t^j$, respectively. They also have an additional memory component for each LSTM cell, $c_t^j$. A visualization of an LSTM unit is shown in \autoref{RNN} (b). The gates are calculated in a similar manner as the GRU unit except for the additional term from the memory component.
\begin{equation} \label{eqL1}
i_t^j = F^j( W_i \mathbf{x}_t + U_i \mathbf{h}_{t-1} + V_i \mathbf{c_{t-1}})
\end{equation}
\begin{equation} \label{eqL2}
o_t^j = F^j( W_o \mathbf{x}_t + U_o \mathbf{h}_{t-1} + V_o \mathbf{c_t})
\end{equation}
\begin{equation} \label{eqL3}
f_t^j = F^j( W_f \mathbf{x}_t + U_f \mathbf{h}_{t-1} + V_j \mathbf{c_{t-1}})
\end{equation}
where $V_i$, $V_o$, and $V_j$ are trainable diagonal matrices. This keeps the memory components internal within each LSTM unit.

The memory component is updated by forgetting the existing content and adding a new memory component $\hat{c}_t^j$:
\begin{equation} \label{eqL4}
c_t^j = f_t^j c_{t-1}^j + i_t^j \hat{c}_t^j
\end{equation}
where the new memory content can be computed by:
\begin{equation} \label{eqL5}
\hat{c}_t^j = G^j(W_c \mathbf{x}_t + U_c \mathbf{h}_{t-1})
\end{equation}
Note how the updated equation for the memory component is governed by the forget and input gates. Finally, the output of the LSTM unit is computed from the memory modulated by the output gate according to the following equation:
\begin{equation} \label{eqL6}
h_t^j = o_t^j tanh(c_t^j)
\end{equation}

Previous works using deep learning with EEG signals have explored the use of CNN-LSTM cascades \cite{Zhang2017}. However, GRUs have been shown in many settings to often match or even beat LSTMs \cite{Chung2014,Tang2016,KIM2017}. GRUs have the ability to perform better with a smaller amount of training data and are faster to train than LSTMs. Thus, in this work, CNN-GRU cascades are also explored and compared against the CNN-LSTM in both accuracy and training speed.

\section{Methodology}
\textcolor{red}{In this section, we first illustrate the DEAP affective EEG dataset \cite{Koelstra2012} that we used to conduct experimental studies and also describe the pre-processing step of our solution. Since DEAP was created for mental state classification purposes, we describe our data partition methodology used to make it more suitable to perform the PI task. Finally, we explain the proposed DL approach and its implementation.}

\subsection{Affective EEG Dataset}
\textcolor{red}{In this study, we performed experiments using DEAP affective EEG dataset which is considered as a standard dataset to perform emotion or affective recognition tasks \cite{deap}. Thirty-two healthy participants participated in the experiment. They were asked to watch affective elicited music videos and score subjective ratings (valence and arousal) for forty video clips during the EEG measurement.}

\begin{table}[t]
\centering
\caption{Affective EEG Data Format with Label}
\label{data}
\begin{tabular}{cc}
\hline
{ \textbf{Array Name}} & { \textbf{Array Shape}}                                                                                 \\ \hline
{ data}                & { \begin{tabular}[c]{@{}c@{}}32 x 40 x 32 x 8064\\ participant x video/trial x EEG x data\end{tabular}} \\ \hline
{ labels}              & { \begin{tabular}[c]{@{}c@{}}32 x 40 x 2\\ participant x video/trial x (valence, arousal)\end{tabular}}                    \\ \hline
\end{tabular}
\end{table}

\textcolor{red}{A summary of the dataset is given in \autoref{data} . The EEG dataset was pre-processed using the following steps:
\begin{itemize}
\item The data was down-sampled to 128 Hz.
\item EOG artifacts were removed using the blind source separation technique called independent component analysis (ICA).
\item A bandpass filter from 4.0--45.0 Hz was applied to the original dataset. The signal was further filtered into different bands as follows: Theta (4--8 Hz), Alpha (8--15 Hz), Beta (15--32 Hz), Gamma (32--40 Hz), and all bands (4--40 Hz).
\item The data was averaged to a common reference.
\item The data was segmented into 60-second trials, and the 3-second pre-trial segments were removed.
\end{itemize}
Most researchers have been using this dataset to develop an affective computing algorithm; however, we used this affective dataset for studying EEG-based PI.}
\subsection{Subsampling and Cross Validation}

Affective EEG is categorised by the standard subjective measures of valence and arousal scores (1--9), with 5 as the threshold for defining low (score $<$ 5) and high (score $\geq$ 5) levels for both valence and arousal. Thus, there were four affective states in total, as stated in \autoref{class}. To simulate practical PI applications, we randomly selected 5 EEG trials per state per person (recorded EEG from 5 video clips) for the experiments. Thus, new users can spend just 5 minutes watching 5 videos for the first registration. \autoref{class} presents a number of subjects in each affective state. The numbers were different in each state because some subjects had less than 5 recorded EEG trials categorized into the state. \textcolor{red}{Furthermore, we aimed to identify a person from a short-length of EEG: 10-seconds. Each EEG trial in DEAP lasts long 60-seconds as a stimulus video. Thus, we simply cut one EEG trial into 6 subsamples. Finally, we had 30 subsamples (6 subsamples $\times$ 5 trials) from each participant in each of the affective states.}

In summary, labels in our experiments (personal identification) are ID of participants. Data and labels that have been used can be described as:
\begin{itemize}{
\item Data: number of participants $\times$ 30 subsamples $\times$ 1280 EEG data points (10-seconds with 128 Hz sampling rate)
\item Label: number of participants $\times$ 30 subsamples  $\times$ 1(ID)}
\end{itemize}
\textcolor{red}{In all experiments, the training, validation, and testing sets were obtained using stratified 10-fold cross-validation. As for the subsamples, 80\% of them were used as training data. As for the validation and test sets, each of them contained 10\% of the subsamples. In each fold, we make sure that the subsamples from each trial is not assigned to more than one set. That is, it can be either in the training, validation or test set. Thus, the subsamples in the training, the validation and the test sets were totally independent.}

\begin{table}[t]
\centering
\caption{Number of participants in each state after subsampling}
\label{class}
\begin{tabular}{ccc}
\hline
{ \textbf{Affective States}}       & { \textbf{Number of Participants}} \\ \hline
{ Low Valence, Low Arousal (LL)}   &  {26}                              \\ \hline
{ Low Valence, High Arousal (LH)}  & {24}                              \\ \hline
{ High Valence, Low Arousal (HL)}  & {23}                              \\ \hline
{ High Valence, High Arousal (HH)} & {32}                              \\ \hline
{ All States}     & {32}                              \\ \hline
\end{tabular}
\end{table}

\subsection{Experiment I: Comparison of affective EEG-based PI among different affective states}
\textcolor{red}{Since datasets contains EEG from five affective states as also shown in \autoref{class}, an experiment was carried out to evaluate which affective states would provide the highest \emph{CRR} in EEG based PI applications. To achieve this goal, two approaches were implemented: deep learning and conventional machine learning. EEG in the range of 4--40 Hz was used in this experiment.}

\subsubsection{Deep Learning Approach}
\begin{figure*}[t]
\centering
\includegraphics[width=0.9\linewidth]{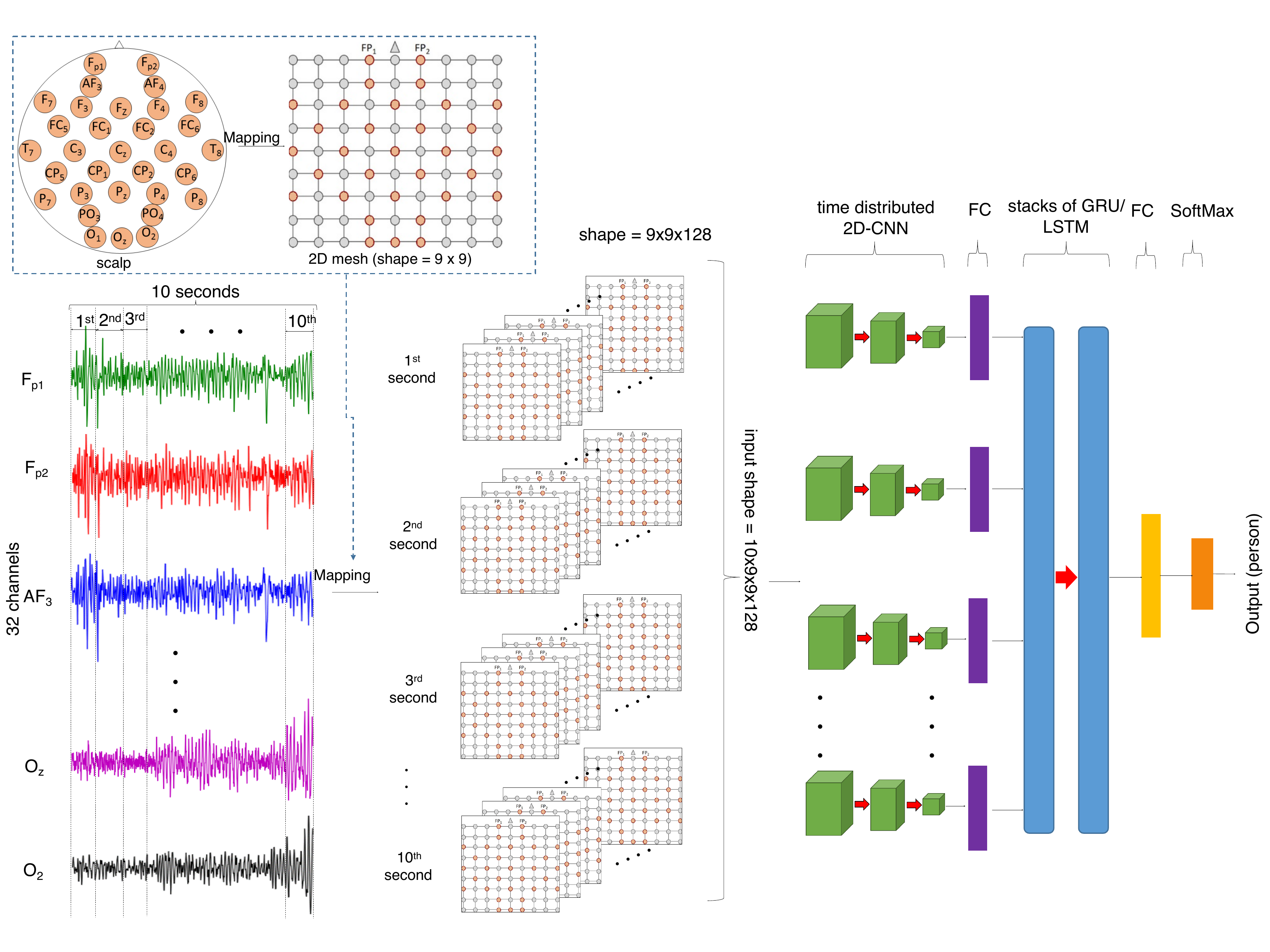}
\caption{Implementation of the cascade CNN-GRU/LSTM model according to EEG data. Meshing is the first step in converting multi-channel EEG signals into sequences of 2D images. The 2D mesh time series is passed through the cascade of CNN and recurrent layers for training, validation, and testing.} 
\label{algor} 
\end{figure*}

\textcolor{red}{\autoref{algor} demonstrates the preparation of the 10-second EEG data before feeding into the DL model. In general, a single EEG channel is a one-dimensional (1D) time series. However, multiple EEG channels can be mapped into time series of 2D mesh (similar to a 2D image). For each time step of the input, the data point from each EEG channel is arranged into one 2D mesh shape of 9$\times$9. 2D mesh size is empirically selected according to the international standard of an electrode placements (10-20 system) with covering all 32 EEG channels. The mesh point (similar to the pixel) which is not allocated for EEG channel is assigned to zeros value throughout the sequences.  The mean and variance for each mesh (32 channels) is normalised individually. In this study, a non-overlapping sliding window is used to separate the data into one-second chunks. Since the sampling rate of input data is 128 Hz, the window size is 128 points. Thus, for each 10-second EEG data, a 10$\times$9$\times$9$\times$128-dimensional tensor is obtained.}

The deep learning model starts with three layers of 2D-CNN (applied to the mesh structure). Each mesh frame from the 128 windows is considered individually in the 2D-CNN. Since this is also a time series, the 2D-CNN is applied to each sliding window, one window at a time, but with shared parameters. This structure is called a TimeDistributed 2DCNN layer. After the TimeDistributed 2DCNN layers, a TimeDistributed Fully Connected (FC) layer is used for subsampling and feature transformation. To capture the temporal structure, two recurrent layers (GRU or LSTM layers) are then applied along the dimension of the sliding windows. Finally, a FC layer is applied to the recurrent output at the final time step with a softmax function for person identification.

The following specific model parameters are used in Experiments I--III. Three layers of TimeDistributed 2DCNNs with 3$\times$3 kernels. We set the number of filters to 128, 64 and 32 for the first, second and third layer respectively. ReLu nonlinearity is used. Batch normalization and dropout are applied after every convolutional layer. For the recurrent layers, we used 2 layers with 32 and 16 recurrent units, respectively. Recurrent dropout was also applied. The dropout rates in each part of the model were fixed at 0.3. We used RMSprop optimizer with a learning rate of 0.003 and a batch size of 256. Although these parameters are held fixed, these settings were found to be good enough for our purposes. The effect of parameter tuning for DL models will be further explored in Experiment IV.

\subsubsection{Conventional Machine Learning Approach using Support Vector Machine (SVM)}

The algorithm aims to locate the optimal decision boundaries for maximising the margin between two classes in the feature space \cite{weston}. This can be done by minimizing the loss:

\begin{equation} \label{eq:erl}
\frac{1}{2}w^tw + C \sum_{i=1}^{n}\xi_i,
\end{equation}
under the constraint 

\begin{equation}
y_i(w^t\phi(x_i)+b)\geq 1-\xi_i  \text{  and  } 
\xi_i\geq0,i=1,...,n.
\end{equation}

$C$ is the capacity constant, $w$ is the vector of coefficients, $b$ is a bias offset, and $y_i$ represents the label of the $i$-th training example from the set of $N$ training examples. The larger the $C$ value, the more the error is penalized. The $C$ value is optimized to avoid overfitting using the validation dataset described earlier.

In the study of person identification, the class label represents the identity number of the participant, considered as a multi-class classification problem. Numerous SVM algorithms can be used such as the ``one-against-one'' approach, ``one-vs-the-rest'' approach \cite{weston}, or k-class SVM \cite{knerr}. To illustrate a strong baseline, the ``one-against-one'' approach, which requires higher computation, is chosen for its robustness towards imbalanced classes and small amounts of data. The ``one-against-one'' SVM solves multi-class classification by building classifiers for all possible pairs of classes resulting in $\frac{N(N-1)}{2}$ classifiers. The predicted class label is the one most yielded from all classifiers.

In this work, the Welch's method is employed as the feature extraction method for the SVM. It is a well-known PSD estimation method, for reducing the variance in periodogram estimation by breaking the data into overlapped segments. Before feeding into the SVM, a normalization step is performed. For normalization, Z-score scaling is adopted, because, experimentally, it performs better than other normalization methods such as min-max and unity normalization in EEG signal processing.

\begin{equation} 
x_{normalized} = \frac{x-\bar{x}_{train}}{s_{train}}
\end{equation} 
Normalization parameters, sample mean$(\bar{x}_{train})$ and sample standard deviation$(s_{train})$, are computed over the training set. The validation set is used to determine the best parameter $C$ chosen from ${0.01,0.1,1,10.100}$ for each experiment.

\textit{Note: according to the results from Experiment I (EX I), DL approaches perform perfectly even when using a mixture of affective state (all states). The affective states do not affect PI performance for DL models. Therefore, the affective EEG states were not considered, and the \emph{all states} setting is used in the remaining experiments.}

\subsection{Experiment II: Comparison of affective EEG-based PI among EEGs from different frequency bands } 
EEG is conventionally used to measure variations in electrical activity across the human scalp. The electrical activity occurs from the oscillation of billions of neural cells inside the human brain. Most researchers usually divided EEG into frequency bands for analysis. \textcolor{red}{Here, we defined Theta (4--8 Hz), Alpha (8--15 Hz), Beta (15--32 Hz), Gamma (32--40 Hz) and all bands (4--40 Hz). Typical Butterworth bandpass filter had incorporated to extract EEGs from different frequency bands.} In this study, we question whether or not frequency bands affect PI performance. To answer the question, we incorporate CNN-LSTM (stratified 10-fold cross-validation), CNN-GRU, and SVM (as performed in EX I) for \emph{CRR} comparison.

\textit{Note: according to the results of Experiment II, all bands (4--40 Hz) provided the best \emph{CRR} and we continued to use all bands for the remainder of the study.}
\subsection{Experiment III: Comparison of affective EEG-based PI among EEGs from sets of sparse EEG electrodes}
\begin{figure} [t]
    \centering
  \subfloat[Frontal (F)]{
       \includegraphics[width=0.41\linewidth]{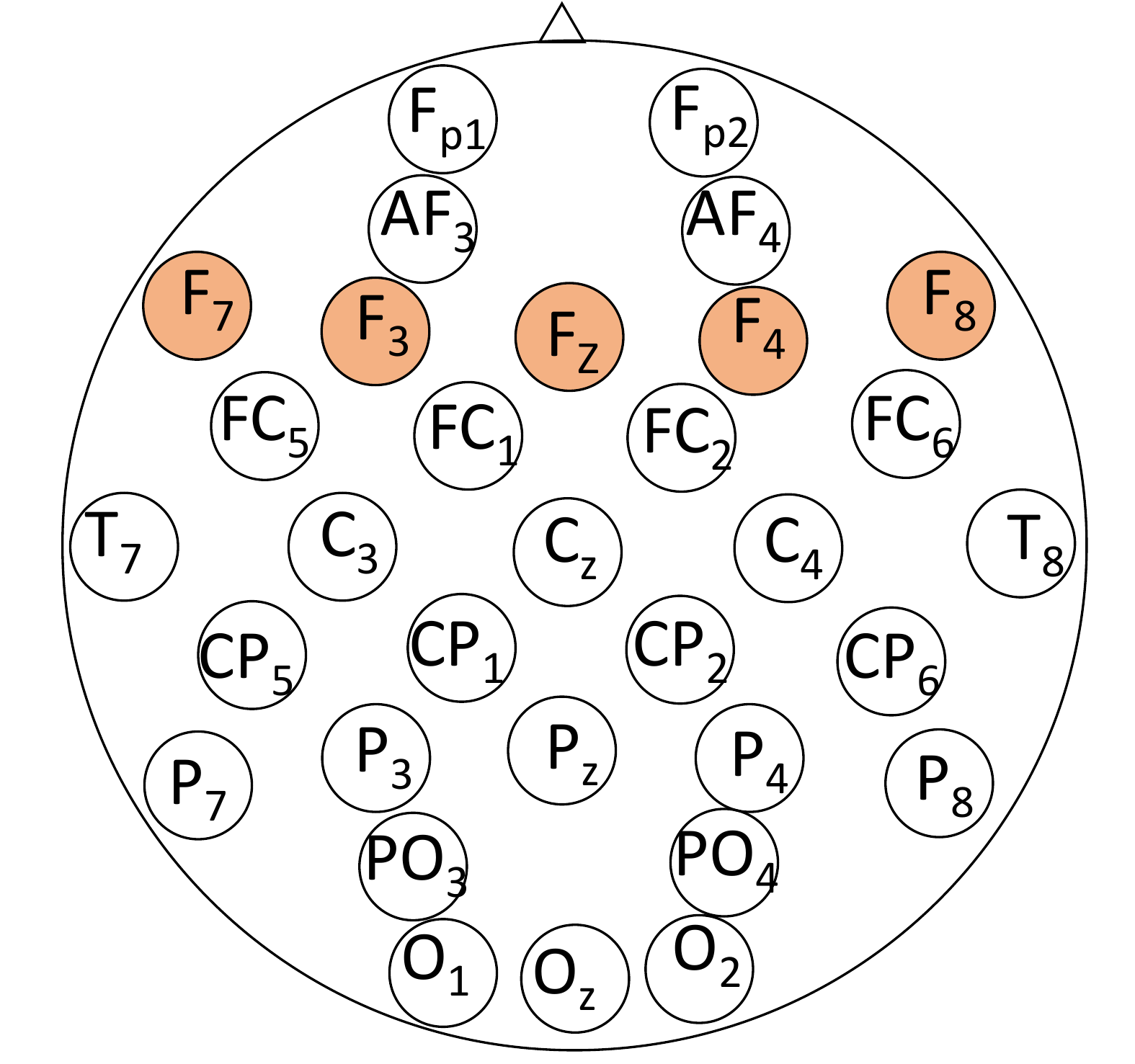}}
    \label{1a}
  \subfloat[Central and Parietal (CP)]{
        \includegraphics[width=0.41\linewidth]{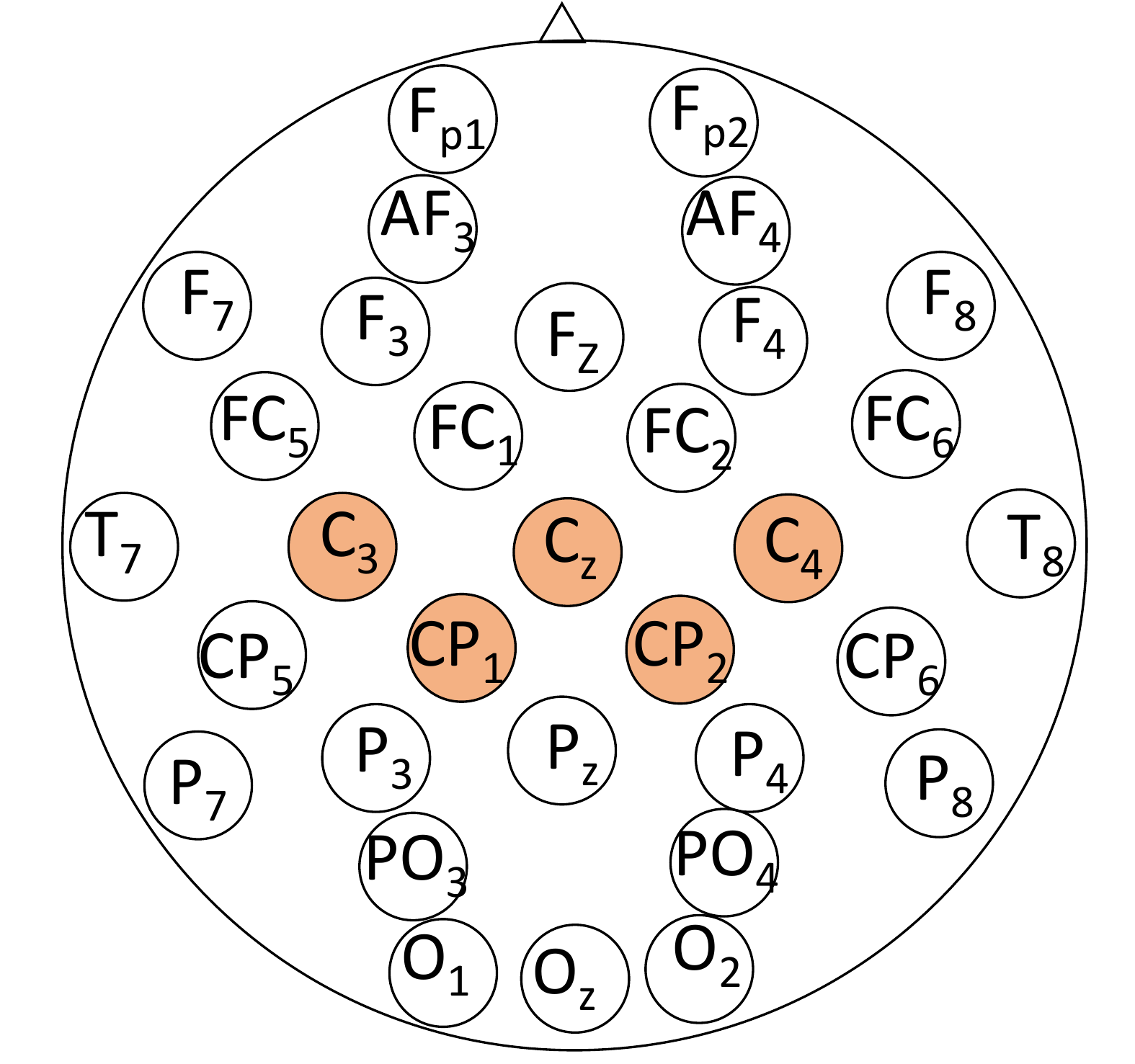}}
    \label{1b}
  \subfloat[Temporal (T)]{
        \includegraphics[width=0.41\linewidth]{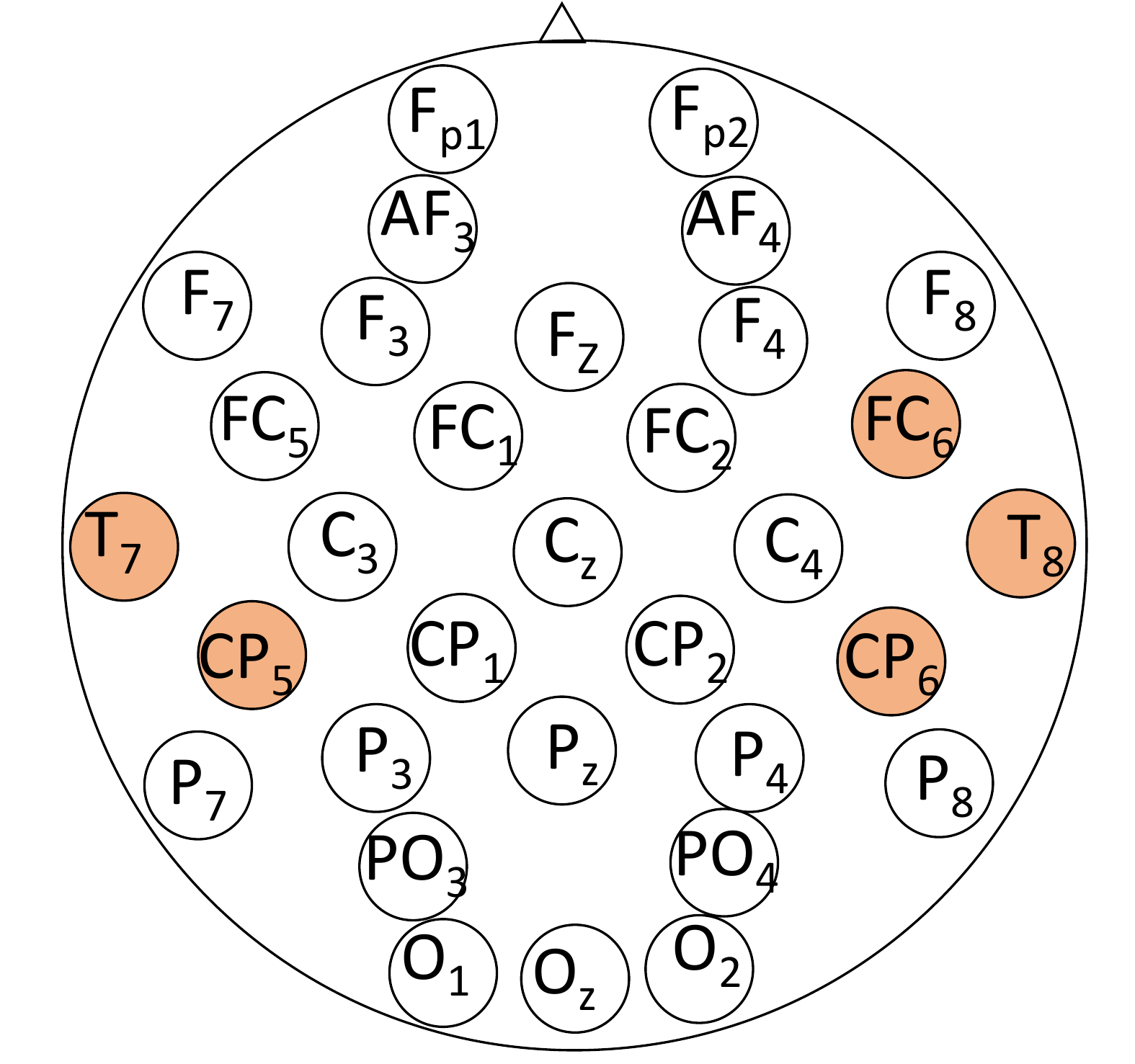}}
    \label{1c}
  \subfloat[Occipital and Parietal (OP)]{
        \includegraphics[width=0.41\linewidth]{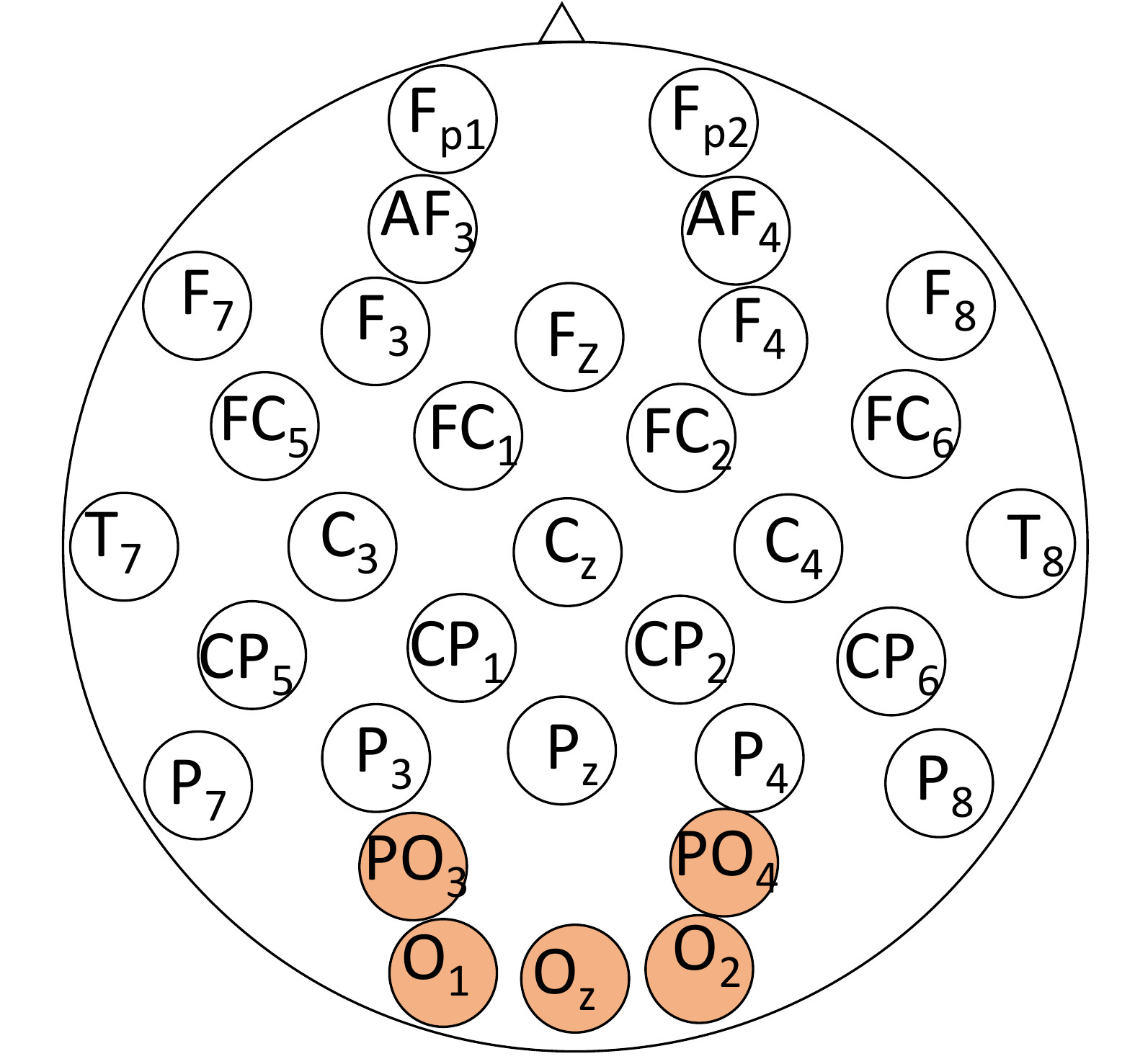}}
     \label{1d} 
    \subfloat[Frontal and Parietal (FP)]{
        \includegraphics[width=0.41\linewidth]{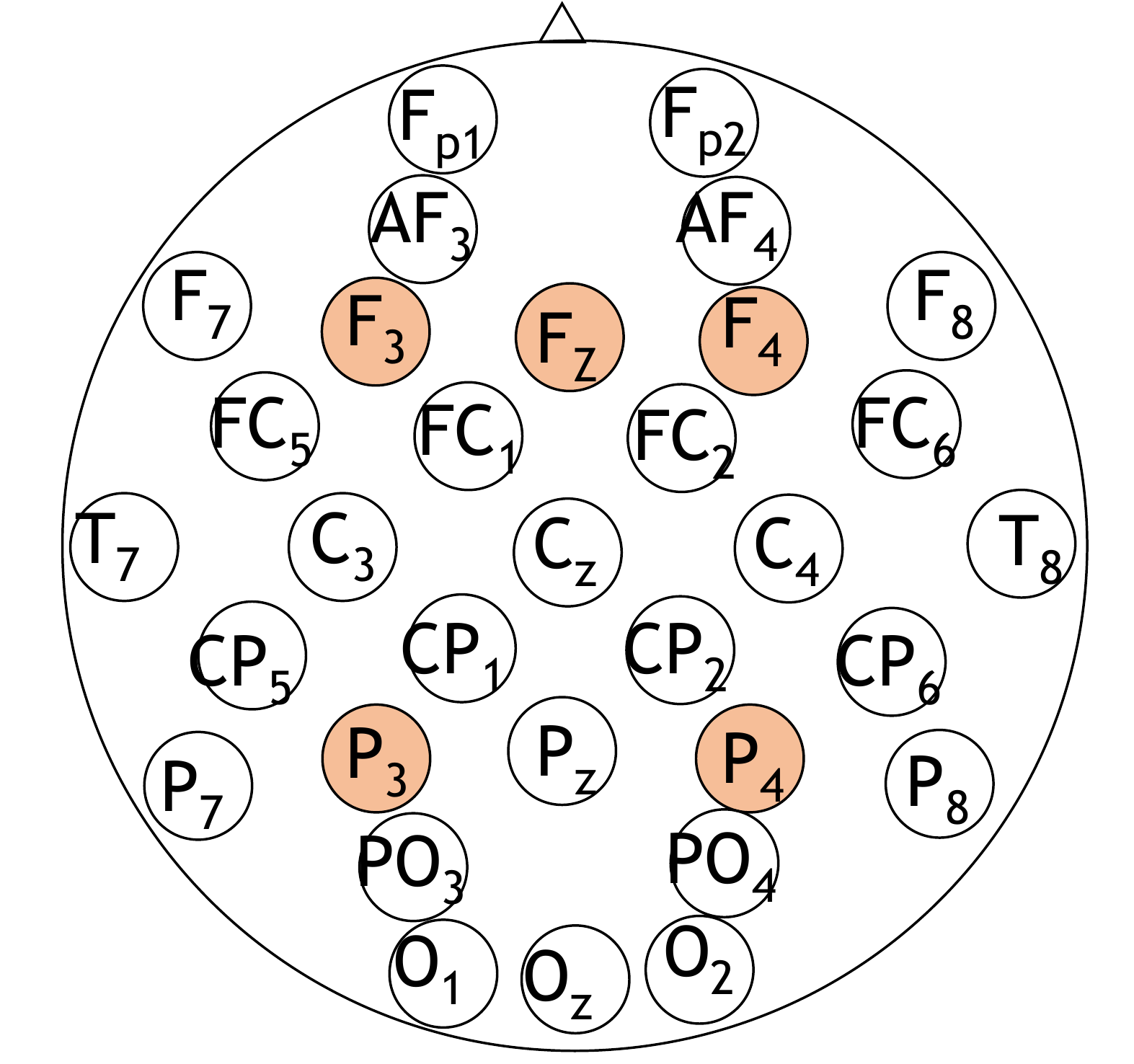}}
     \label{1e} 
  \caption{Experimental Study III evaluates the \emph{CRR} of the EEG-based PI in different sets of sparse EEG electrodes. Five EEG electrode channels from each part of the scalp were grouped into five different configurations (a-e)}
  \label{brain} 
\end{figure}

In this experiment, we hypothesized whether or not the number of electrodes could be reduced from thirty-two channels to five while maintaining an acceptable \emph{CRR}. The lower the number of electrodes required, the more user-friendly and practical the system. To investigate this question, we defined sets of five EEG electrodes as shown in \autoref{brain}, including Frontal (F) \autoref{brain}(a), Central and Parietal (CP) \autoref{brain}(b), Temporal (T) \autoref{brain}(c), Occipital and Parietal (OP) \autoref{brain}(d), and Frontal and Parietal (FP) \autoref{brain}(e). According to EX I and II, the DL approach significantly outperforms the traditional SVM in PI applications. Thus, we incorporated only CNN-GRU and CNN-LSTM in this investigation.

\begin{table}[]
\centering
\caption{Variation in the number of filters for each CNN layers while fixing the number of GRU/LSTM units}
\label{CNN}
\begin{tabular}{ccc}
\hline
{ \textbf{CNN}} & { \textbf{GRU/LSTM}} \\ \hline
{128}          & {32. 16}            \\ \hline
{128, 64}      & {32, 16}            \\ \hline
{128, 64, 32}  & {32, 16}            \\ \hline
\end{tabular}
\end{table}

\begin{table}[]
\centering
\caption{Variation in the number of GRU/LSTM units while fixing the CNN layers}
\label{GRU}
\begin{tabular}{ccc}
\hline
{ \textbf{CNN}} & { \textbf{GRU/LSTM}} \\ \hline
{ 128, 64, 32}  & { 16, 8}             \\ \hline
{ 128, 64, 32}  & { 32, 16}            \\ \hline
{ 128, 64, 32}  & { 64, 32}            \\ \hline
\end{tabular}
\end{table}

\subsection{Experiment IV: Comparison of proposed CNN-GRU against CNN-LSTM and other relevant approaches towards affective EEG-based PI application}
First, we evaluated our proposed CNN-GRU against a spatiotemporal DL model, namely which CNN-LSTM \cite{Zhang2017}. Both approaches have been previously described in detail in Section III and \autoref{algor}. In this study, we measured the performance in terms of the mean \emph{CRR} and the convergence speed as we tuned the size of the models by varying the number of CNN layers and the number of GRU/LSTM units.
We also compared our best models against other conventional machine learning methods and relevant works, such as Mahalanobis distance-based classifiers, using either PSD or spectral coherence (COH) as features (reproduced from \cite{LaRocca2014}) and DNN/SVM as proposed in \cite{li2017}.

\subsubsection{Deep Learning Approach}
To find suitable CNN layers for cascading with either GRU or LSTM, the numbers of CNN layers were varied as presented in \autoref{CNN}. The selected CNN layers were then cascaded to GRU/LSTM and the numbers of GRU and LSTM units varied, as can be seen in \autoref{GRU}.

\subsubsection{Baseline Approach}
As previously mentioned, the Mahalanobis distance-based classifier with either PSD or COH as features was used as a baseline. This approach was reported to provide the highest \emph{CRR} among multiple approaches in a recent critical review paper on EEG-based PI \cite{Yang2017}. However, it has never been applied on the DEAP affective datasets.

To obtain the PSD and COH features, the same parameters were used as reported in \cite{LaRocca2014}, except that the number of FFT points was set to 128. Each PSD feature has $N_{PSD} = 32$ elements (electrodes) and each COH feature has $N_{COH} = 496$ elements (pairs). Classification was then performed on the transformed features. Fisher's Z transformation was applied to the COH features and a logarithmic function to the PSD features. After the transformed PSD and COH features for each element were obtained, the Mahalanobis distances, $d_{m,n}$, were then computed as shown in Equation \ref{eq:mahalanobis}.

\begin{equation}
d_{m,n} = (O_m - \mu_n)\Sigma^{-1}(O_m - \mu_n)^T
\label{eq:mahalanobis}
\end{equation}

where $O_m$ is the observed feature vector, $\mu_n$ is the mean feature vector of class $n$, and $\Sigma^{-1}$ is the inverse pooled covariance matrix. The pooled covariance matrix is the averaged-unbiased covariance matrix of all class distributions. For each sample, the Mahalanobis distances were computed between the observed sample m and the class distribution n, thus a distance vector of size N where N = 32, representing the number of classes (participants) in the dataset.

Two different schemes were used in \cite{LaRocca2014}. The first scheme was a single-element classification to perform the identification of each electrode separately. The other scheme was the all-element classification, combining the best subset of electrodes using match score fusion. We chose the all-element classification scheme which yielded better performance. We modified the scheme to be compatible to this work by selecting all electrodes instead of choosing just a subset. Stratified 10-fold cross-validation was also performed on the all-element classification to obtain the mean \emph{CRR}.

\section{Results}
Experimental results are reported separately in each study. Then all of them are summarized at the end of the section.
\subsection{Results I: comparison of affective EEG-based PI among different affective states} 
The comparison of the mean correct recognition rate or \emph{CRR} (with standard error bar) among different affective states and different recognized approaches had been shown in \autoref{G1}. Statistical testing named one way repeated measures ANOVA (no violation on Sphericity Assumed) with Bonferroni pairwise comparison (post-hoc comparison) had been implemented for comparison of mean \emph{CRR} (stratified 10-fold cross-validation).

\begin{table}[t]
\centering
\caption{Comparison of the mean correct recognition rate or \emph{CRR} (with standard error bar) among different affective states and different recognised approaches. CNN-GRU and CNN-LSTM significantly outperformed the traditional SVM in every affective state (including all states), * notes \emph{p$<$0.01}. EEG in the range of 4--40 Hz has been used in this experiment.}
\label{G1}
\begin{tabular}{cccc}
\hline
{ \textbf{}}          & \multicolumn{3}{c}{{ \textbf{Mean CRR {[}\%{]}}}}                                                    \\ \hline
{ \textbf{States}}           & { \textbf{CNN-GRU}} & { \textbf{CNN-LSTM}} & { \textbf{SVM}}   \\ \hline
{ \textbf{LL}}         & { \textbf{99.90  $\pm$ 0.10*}} & { 99.79 $\pm$ 0.14*}    & { 33.02 $\pm$ 1.58} \\ \hline
{ \textbf{LH}}         & { 99.71 $\pm$ 0.19*}  & { \textbf{100.0*}}    & { 36.38 $\pm$ 1.71} \\ \hline
{ \textbf{HL}}         & { \textbf{99.86 $\pm$ 0.14*}}  & { \textbf{99.86 $\pm$ 0.14*}}    & { 36.25 $\pm$ 2.45} \\ \hline
{ \textbf{HH}}         & { \textbf{99.87 $\pm$ 0.12*}}   & { 99.74 $\pm$ 0.26*}    & { 33.59 $\pm$ 1.65} \\ \hline
{ \textbf{All States}} & { \textbf{100.0*}} & { 99.79 $\pm$ 0.14*}   & { 33.02 $\pm$ 1.58} \\ \hline
\end{tabular}
\end{table}

In the comparison of \emph{CRR} among different affective states, the statistical results demonstrate that the \textit{EEG (4--40 Hz) from different affective states does not affect the performance of affective EEG-based PI in all recognised approaches (F(4)=0.805, p$=$0.530}, \emph{F(4)=0.762, p$=$0.557} and \emph{F(4)=0.930, p$=$0.457} for CNN-GRU, CNN-LSTM, and SVM, respectively).

Moreover, in comparison of \emph{CRR} among different approaches, the statistical results show a significant difference in the mean \emph{CRR} among CNN-GRU, CNN-LSTM, and SVM approaches. In pairwise comparison, CNN-GRU and CNN-LSTM significantly outperformed the traditional SVM in every affective state (including all states), \emph{p$<$0.01}. Both CNN-GRU (in all states) and CNN-LSTM (in LH) reached up to 100\% in mean \emph{CRR}. Further reports on comparative studies of CNN-GRU and CNN-LSTM for EEG-based PI against previous works can be seen in Results IV.

\subsection{Results II: comparison of affective EEG-based PI among EEGs from different frequency bands}
Here, we report the comparison of the mean correct recognition rate or \emph{CRR} among different EEG frequency bands and different recognized approaches (shown in \autoref{G2}) using the same statistical testing as same as in Section IV A).


\begin{table}[t]
\centering
\caption{Comparison of the mean correct recognition rate or CRR (with standard error bar) among different EEG frequency bands and different recognised approaches, no differences in CNN-GRU, CNN-LSTM, and SVM were found in low frequency bands (Theta (4--8 Hz) and Alpha (8--15 Hz)). However, they significantly outperformed the SVM in Beta (15--32 Hz), Gamma (32--40 Hz), and all bands (4--40 Hz), * notes \emph{p$<$0.01}.}
\label{G2}
\begin{tabular}{cccc}
\hline
{ \textbf{}}          & \multicolumn{3}{c}{{ \textbf{Mean CRR {[}\%{]}}}}                                                    \\ \hline
{ }                   & { \textbf{CNN-GRU}} & { \textbf{CNN-LSTM}} & { \textbf{SVM}} \\ \hline
{ \textbf{4-8 Hz}}    & { \textbf{99.69 $\pm$ 0.22}}            & { 99.69 $\pm$ 0.22}             & { 98.54 $\pm$ 0.35}        \\ \hline
{ \textbf{8-15 Hz}}   & { 99.58 $\pm$ 0.23}            & { \textbf{99.69 $\pm$ 0.22}}             & { 98.75 $\pm$ 0.34}        \\ \hline
{ \textbf{15-32 Hz}}  & { \textbf{99.90 $\pm$ 0.10*}}            & { 99.86 $\pm$ 0.16*}             & { 87.50 $\pm$ 0.64}        \\ \hline
{ \textbf{32-40 Hz}}  & { \textbf{100.0*}}            & { 99.74 $\pm$ 0.14*}             & { 33.54 $\pm$ 1.57}        \\ \hline
{ \textbf{all bands}} & { \textbf{100.0*}}            & { 99.79 $\pm$ 0.14*}             & { 33.02 $\pm$ 1.58}        \\ \hline
\end{tabular}
\end{table}

In the comparison of \emph{CRR} among different frequency bands, the statistical results demonstrate that \textit{EEG from different frequency bands does not affect the performance of affective EEG-based PI in CNN-GRU and CNN-LSTM approaches} (\emph{F(4)=2.168, p$=$0.092} and \emph{F(4)=0.144, p$=$0.964} for CNN-GRU and CNN-LSTM, respectively). However, the SVM approach shows that Theta (4--8 Hz) and Alpha (8--15 Hz) provide significantly higher \emph{CRR} than Beta (15--32 Hz), Gamma (32--40 Hz), and all bands (4--40 Hz) (\emph{F(4)=1309.747, p$<$0.01} in ANOVA testing and \emph{p$<$0.01} in all pairwise comparisons).

Furthermore, in the comparison of \emph{CRR} among different approaches, there were no differences in CNN-GRU, CNN-LSTM, and SVM for low frequency bands (Theta (4--8 Hz) and Alpha (8--15 Hz)). However, CNN-GRU and CNN-LSTM significantly outperformed the SVM in Beta (15--32 Hz), Gamma (32--40 Hz), and all bands (4--40 Hz), \emph{p$<$0.01}. CNN-GRU and CNN-LSTM reached up to 100\% and 99.79\%, respectively, in all bands.

\subsection{Results III: Comparison of affective EEG-based PI among EEGs from sets of sparse EEG electrodes} 
\begin{figure}[]
\centering
\includegraphics[width=0.95\linewidth]{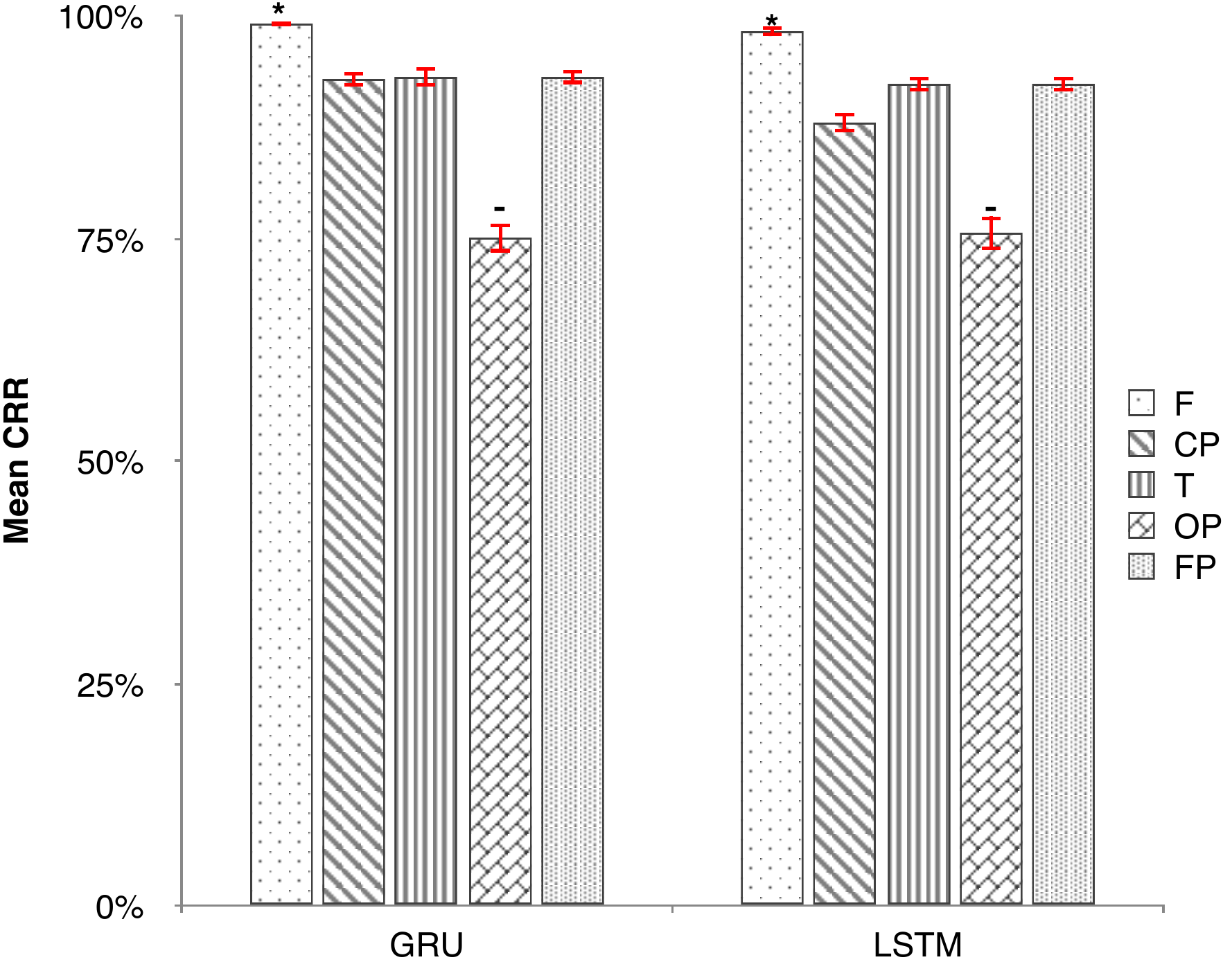}
\caption{Comparison of CRR among five sets of electrodes. The frontal part (F) provided significantly higher \emph{CRR} compared to the others \emph{p$<$0.05}. On the other hand, occipital and parietal (OP) provided significantly lower \emph{CRR} compared to the others \emph{p$<$0.05}.} 
\label{G3} 
\end{figure}

According to \autoref{G3}, one-way repeated measures ANOVA with Bonferroni pairwise comparison (post-hoc) reported that five electrodes in the F set provided a significantly higher mean \emph{CRR} than the other sets in both CNN-GRU and CNN-LSTM \emph{p$<$0.05}. CNN-GRU and CNN-LSTM reached up to (99.17 $\pm$ 0.34\%) and (98.23 $\pm$ 0.52\%) mean \emph{CRR}, respectively (stratified 10-fold cross-validation). To reduce the number of EEG electrodes from thirty-two to five for more practical application, $F_{3}$, $F_{4}$, $F_{z}$, $F_{7}$ and $F_{8}$ were the best five electrodes for application in similar scenarios to this experiment.

\subsection{Results IV: Comparison of proposed CNN-GRU against CNN-LSTM and the other relevant approaches towards affective EEG-based PI application}
\begin{figure}[]
\centering
\includegraphics[width=0.9\linewidth]{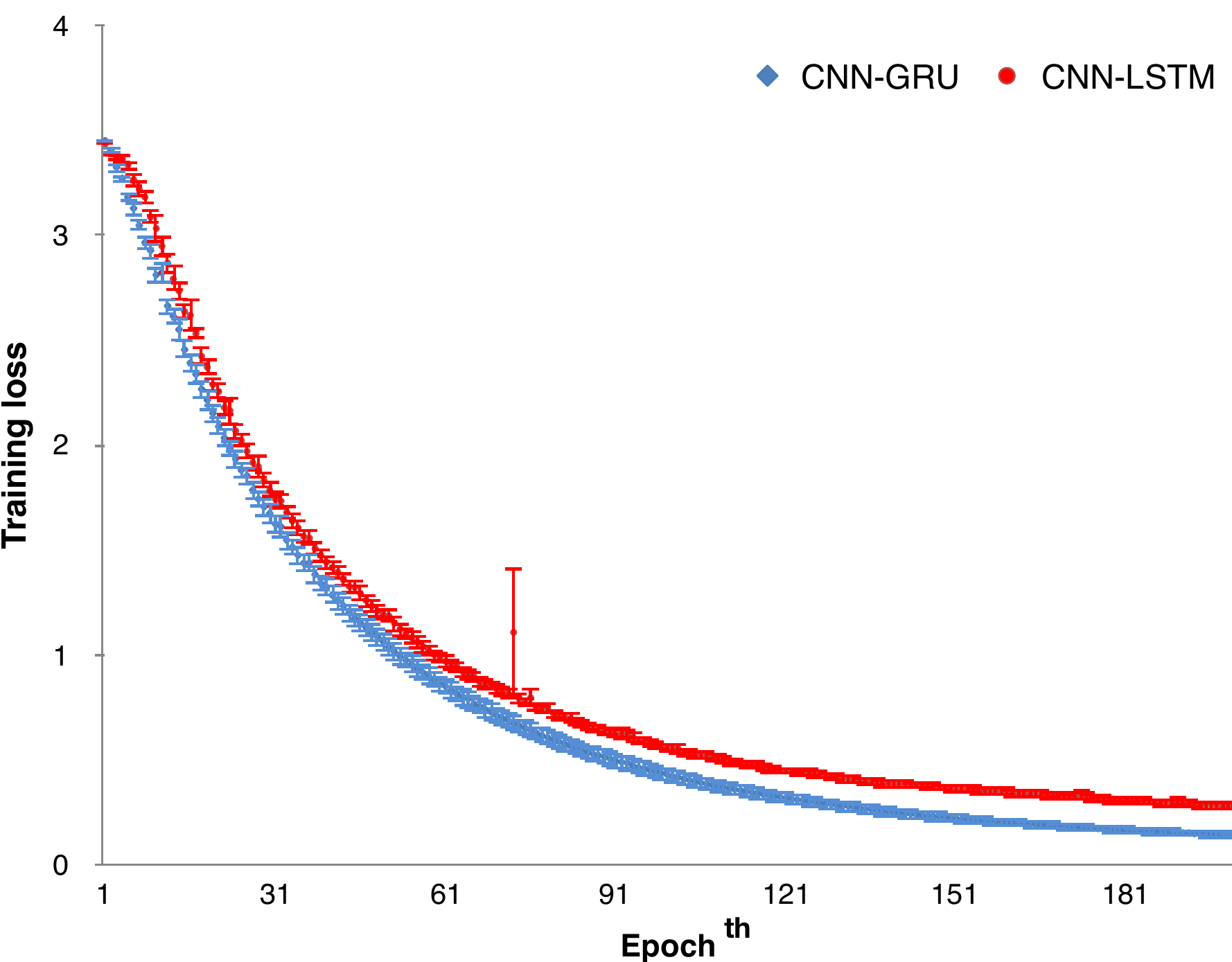}
\caption{Comparison of training loss by epoch between CNN-GRU and CNN-LSTM. The configuration consists of 3 CNN layers with 128, 64, 32 filters and 2 layers of GRU/LSTM with 16 and 8 units.} 
\label{G4} 
\end{figure}

\begin{figure}[]
\centering
\includegraphics[width=0.9\linewidth]{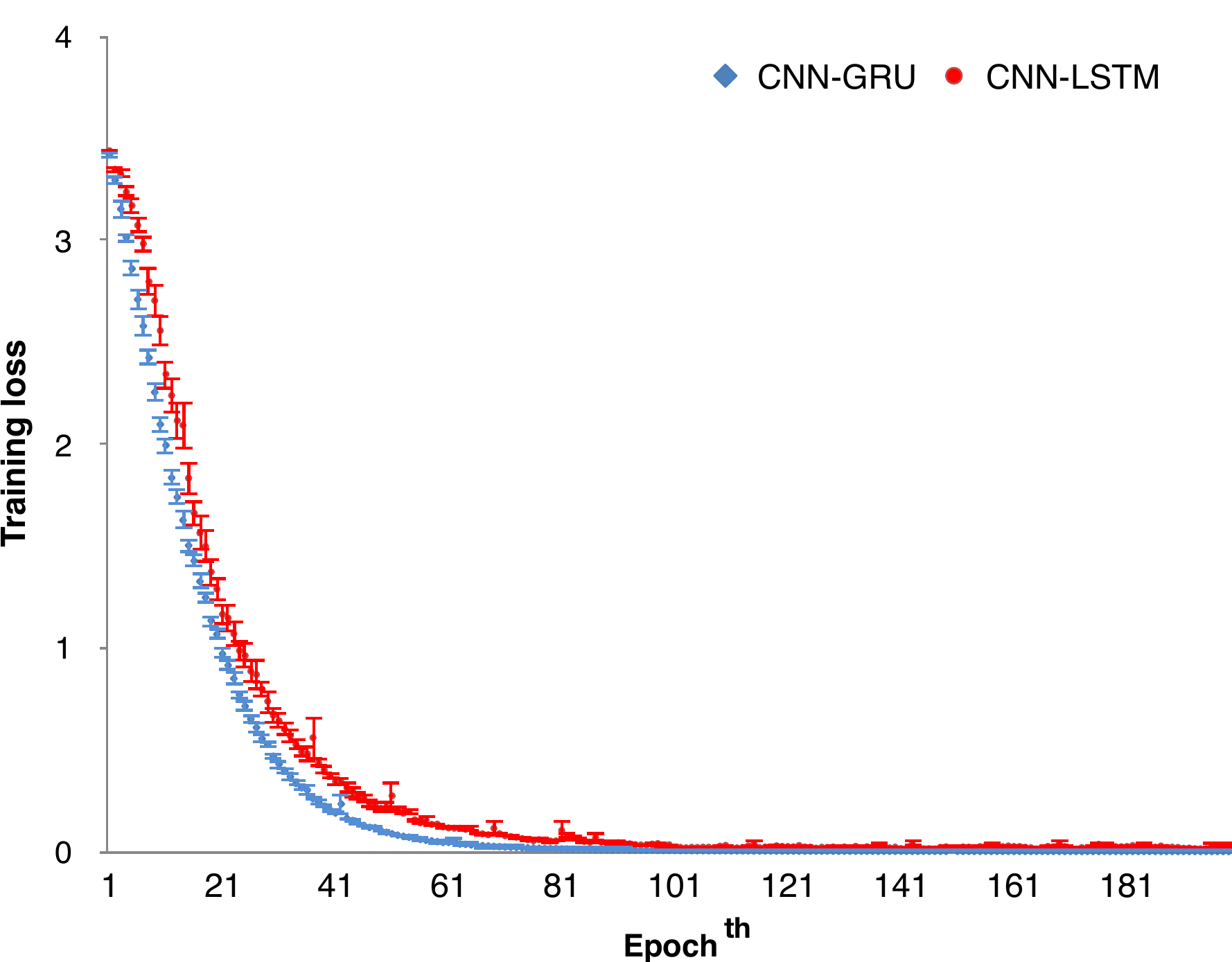}
\caption{Comparison of training loss by epoch between CNN-GRU and CNN-LSTM. The configuration consists of 3 CNN layers with 128, 64, 32 filters and 2 layers of GRU/LSTM with 32 and 16 units.} 
\label{G5} 
\end{figure}

\autoref{DCNN} and \autoref{DRNN} present mean \emph{CRR} (stratified 10-fold cross-validation) from the proposed CNN-GRU against the conventional spatiotemporal DL model, namely CNN-LSTM, in various parameter settings (number of CNN layers and GRU/LSTM units). Mean \emph{CRR}s from CNN-GRU were higher or equal compared to those from CNN-LSTM in all settings. The standard t-test indicated that the mean \emph{CRR} from CNN-GRU was significantly higher (\emph{p$<$0.01}) than that of the CNN-LSTM in 3 CNN layers with 128, 64, and 32 filters and 2 layers of GRU/LSTM with 16 and 8 units. In the comparison of training speed between the two approaches, CNN layers were fixed with 128, 64, 32 filters because the mean \emph{CRR} was equal as shown in \autoref{DCNN}. \autoref{G4} and \autoref{G5} present a training speed comparison (in terms of training loss by epoch) from CNN-GRU and CNN-LSTM on 2 layers of GRU/LSTM with 16, 8 and 32, 16 units, respectively. It was obvious that training loss from CNN-GRU was decreasing faster than from CNN-LSTM. These results were also consistent with GRU/LSTM for 64, 32 units.

\autoref{final} demonstrates EEG-based PI performance using the proposed approach (CNN-GRU) against conventional DL (CNN-LSTM) and the baseline approach. \textcolor{red}{The baseline approaches are reproducing Mahalanobis distance-based classifier using PSD/COH as features and typical non-linear classifiers (DNN or ANN or SVM) from previous works \cite{li2017,Banos}, with the same datasets.} CNN-GRU/LSTM (constructed using CNN layers with 128, 64, 32 filters and 2 layers of GRU/LSTM with 32 and 16 units) produced a higher mean \emph{CRR} than the others. Furthermore, the CNN-GRU was better than CNN-LSTM both in terms of training speed (as shown \autoref{G4}) and mean \emph{CRR} with a small number of electrodes (five from the frontal area).

\begin{table}[]
\centering
\caption{Comparison of mean \emph{CRR} results between different CNN layers with 32 EEG electrodes}
\label{DCNN}
\begin{tabular}{cccc}
\hline
{ }                               & { }                                    & \multicolumn{2}{c}{{ \textbf{Mean CRR {[}\%{]}}}}             \\ \cline{3-4} 
\multirow{-2}{*}{{ \textbf{CNN}}} & \multirow{-2}{*}{{ \textbf{GRU/LSTM}}} & { \textbf{GRU}}        & { \textbf{LSTM}} \\ \hline
{ 128}                            & { 32, 16}                              & { \textbf{100 $\pm$ 0.00}}   & { 99.69 $\pm$ 0.16}    \\ \hline
{ 128, 64}                        & { 32, 16}                              & { \textbf{99.90 $\pm$ 0.10}} & { 99.69 $\pm$ 0.16}    \\ \hline
{ 128, 64, 32}                    & { 32, 16}                              & { 99.90 $\pm$ 0.10}          & { 99.90 $\pm$ 0.10}    \\ \hline
\end{tabular}
\end{table}

\begin{table}[]
\centering
\caption{Comparison of mean \emph{CRR} results between different GRU/LSTM units with 32 EEG electrodes. * denotes that the mean \emph{CRR} is significantly higher, \emph{p$<$0.01}.}
\label{DRNN}
\begin{tabular}{cccc}
\hline
{ }                               & { }                                    & \multicolumn{2}{c}{{ \textbf{Mean CRR {[}\%{]}}}}          \\ \cline{3-4} 
\multirow{-2}{*}{{ \textbf{CNN}}} & \multirow{-2}{*}{{ \textbf{GRU/LSTM}}} & { \textbf{GRU}}                 & { \textbf{LSTM}}       \\ \hline
{ 128, 64, 32}                    & { 16, 8}                               & { \textbf{97.29 $\pm$ 0.75 *}} & { 89.58 $\pm$ 1.81} \\ \hline
{ 128, 64, 32}                    & { 32, 16}                              & { 99.90 $\pm$ 0.10}          & { 99.90 $\pm$ 0.10} \\ \hline
{ 128, 64, 32}                    & { 64, 32}                              & { \textbf{99.90 $\pm$ 0.10}} & { 99.79 $\pm$ 0.25} \\ \hline
\end{tabular}
\end{table}

\begin{table}[]
\centering
\caption{Comparison of mean \emph{CRR} with proposed DL approach (CNN-GRU), conventional DL approach CNN-LSTM, conventional machine learning and previous works on the same datasets.}
\label{final}
\begin{tabular}{ccc}
\hline
{ \textbf{Approach}} & { \textbf{Numbers of Electrodes}} & { \textbf{Mean CRR {[}\%{]}}} \\ \hline
{ CNN-GRU}           & { 32}                             & { \textbf{99.90 $\pm$ 0.11}}        \\ \hline
{ CNN-GRU}           & { \textbf{5}}                     & { 99.10 $\pm$ 0.34}                 \\ \hline
{ CNN-LSTM}          & { 32}                             & { \textbf{99.90 $\pm$ 0.11}}        \\ \hline
{ CNN-LSTM}          & { 5}                              & { 98.23 $\pm$ 0.52}                 \\ \hline
{ Mahalanobis-PSD}   & { 32}                             & { 47.09 $\pm$ 2.34}             \\ \hline
{ Mahalanobis-COH}   & { 32}                             & { 47.81 $\pm$ 3.29}                           \\ \hline
{ DNN \cite{li2017}}        & { 8}                              & { 85.0 $\pm$ 4.0}                   \\ \hline
{ SVM \cite{li2017}}        & { 8}                              & { 88.0 $\pm$ 4.0}                   \\ \hline
\textcolor{red}{ SVM or ANN-PSD \cite{Banos}}        & {32}                              & {97.97}                   \\ \hline
\end{tabular}
\end{table}

\section{Discussion}
From the experimental results, we will focus on two kinds of issues, namely physical and algorithmic issues for affective EEG-based PI applications. The physical issues refer to the EEG capturing such as the different affective states, the frequency bands, and the electrode positions on the scalp. The algorithmic issues were about how to use the proposed approach (CNN-GRU) on EEG in an effective way for PI applications and the advantages of CNN-GRU over the other relevant approaches.

Regarding the physical issues, the experimental results indicate that DL approaches (CNN-GRU and CNN-LSTM) can deal with EEG (4--40 Hz) from different affective states (valence and arousal levels), reaching up to 100\% mean \emph{CRR}. On the other hand, a traditional machine learning approach such as SVM using PSD as features did not reach 50\% mean \emph{CRR}. \textcolor{red}{However, the SVM approach was found to improve considerably when focusing on specific EEG frequency bands, namely Theta (4--8 Hz) and Alpha (8--15 Hz). The SVM reached up to 98\% mean \emph{CRR} with either Theta or Alpha EEG. As for CNN-GRU and CNN-LSTM, EEG frequency bands had little to no effect because that the DL approaches can capture various hidden features (including non-frequency related features). Thus the hidden features can still maintain high percentages of \emph{CRR}} To reduce the number of EEG electrodes from thirty-two to five for more practical applications, $F_{3}$, $F_{4}$, $F_{z}$, $F_{7}$ and $F_{8}$ were the best five electrodes. CNN-GRU and CNN-LSTM reached up to 99.17\% and 98.23\% mean \emph{CRR}, respectively. The results show that EEG electrodes from the frontal scalp provided higher mean \emph{CRR} than other positions on the scalp, which is consistent with previous work on EEG-based PI \cite{LaRocca2014}.

Concerning the algorithmic issue, the proposed CNN-GRU and conventional spatiotemporal DL models (CNN-LSTM) for EEG-based PI outperformed the state-of-the-art and relevant algorithms (Mahalanobis distance-based classifier using PSD/COH as features, DNN, and SVM \cite{LaRocca2014}\cite{li2017}) on the same dataset. \textcolor{red}{In the comparison between CNN-GRU and CNN-LSTM, CNN-GRU was obviously better in terms of training speed while having a slightly higher mean \emph{CRR}, especially in when using a small amount of electrode. Furthermore, CNN-GRU overcomes the influence of affective states in EEG-Based PI reported in the previous works \cite{Banos, Arnau}.}

\section{Conclusion}
In conclusion, we explored the feasibility of using affective EEG for person identification. We proposed a DL approach called CNN-GRU, as the classification algorithm. EEG-based PI using CNN-GRU reached up to 99.90--100\% mean \emph{CRR} with 32 electrodes, and 99.17\% with 5 electrodes. CNN-GRU significantly outperformed the state-of-the-art and relevant algorithms in our experiments. In the comparison between CNN-GRU and the conventional DL cascade model (CNN-LSTM), CNN-GRU was obviously better in terms of training speed. The mean \emph{CRR} from CNN-GRU was slightly higher than from CNN-LSTM, especially when using only five electrodes. \textcolor{red}{Furthermore, CNN-GRU overcomes the influence of affective states in EEG-Based PI reported in the previous works.}

\bibliographystyle{IEEEtran}
\bibliography{ref2} 
\vskip -2pt plus -1fil
\begin{IEEEbiography}[{\includegraphics[width=1.1in,height=1.4in,clip,keepaspectratio]{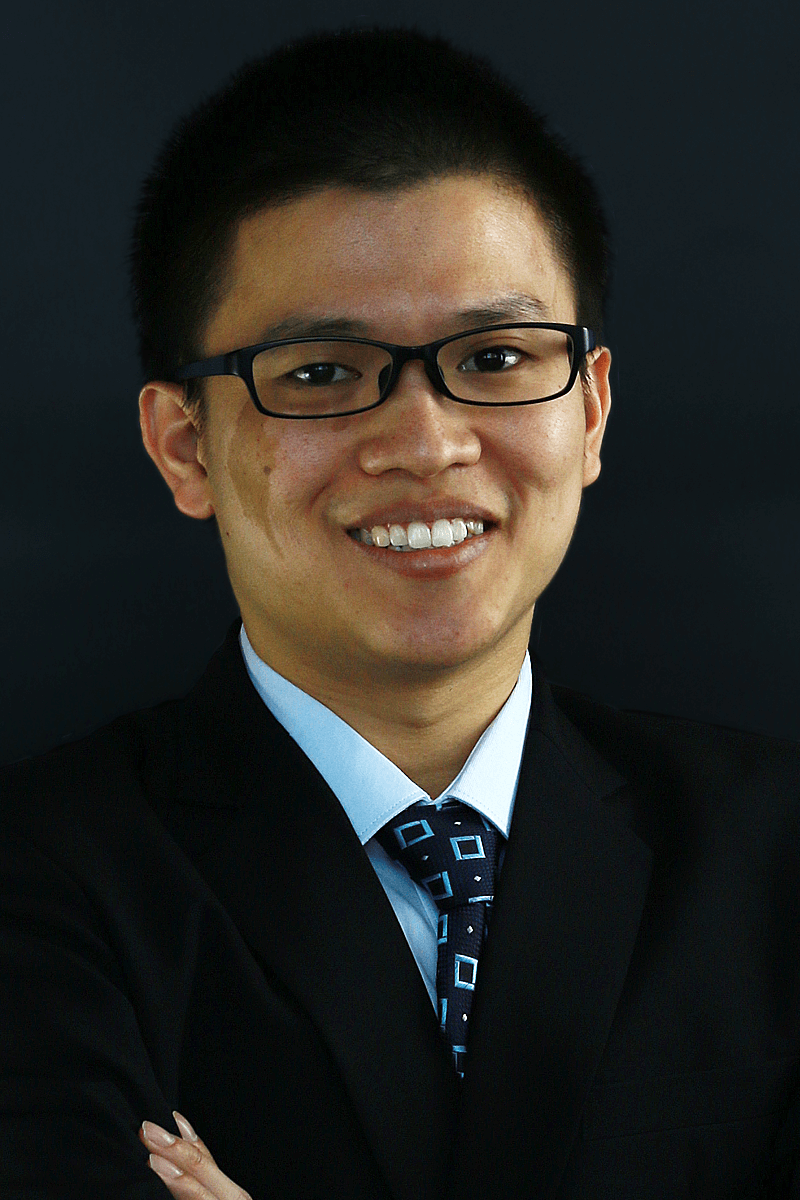}}]{Theerawit Wiaiprasitporn}
received the Ph.D. Degree in Engineering from Graduate School of Information Science and Engineering, Tokyo Institute of Technology, Japan, in 2017. 
Now, he is working as lecturer position at School of Information Science and Technology at Vidyasirimedhi Institute of Science and Technology (VISTEC), Thailand. He is also a co-PI of Bio-inspired Robotics and Neural engineering (BRAIN) lab at VISTEC. His current research are Neural Engineering (BCI), Bio-Potential Applications, Biomedical and Health Informatics and Smart Living.
\end{IEEEbiography}
\vskip -2pt plus -1fil
\begin{IEEEbiography}[{\includegraphics[width=1.1in,height=1.7in,clip,keepaspectratio]{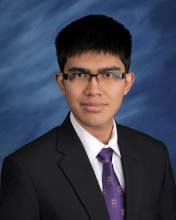}}]
{Apiwat Ditthapron} is currently pursuing B.S. degree from the department of computer science, Worcester Polytechnic Institute, MA, USA. His current research interests include Computer Vision, Machine learning, Deep learning, and Data Visualization.
\end{IEEEbiography}
\vskip -2pt plus -1fil
\begin{IEEEbiography}[{\includegraphics[width=1.1in,height=1.7in,clip,keepaspectratio]{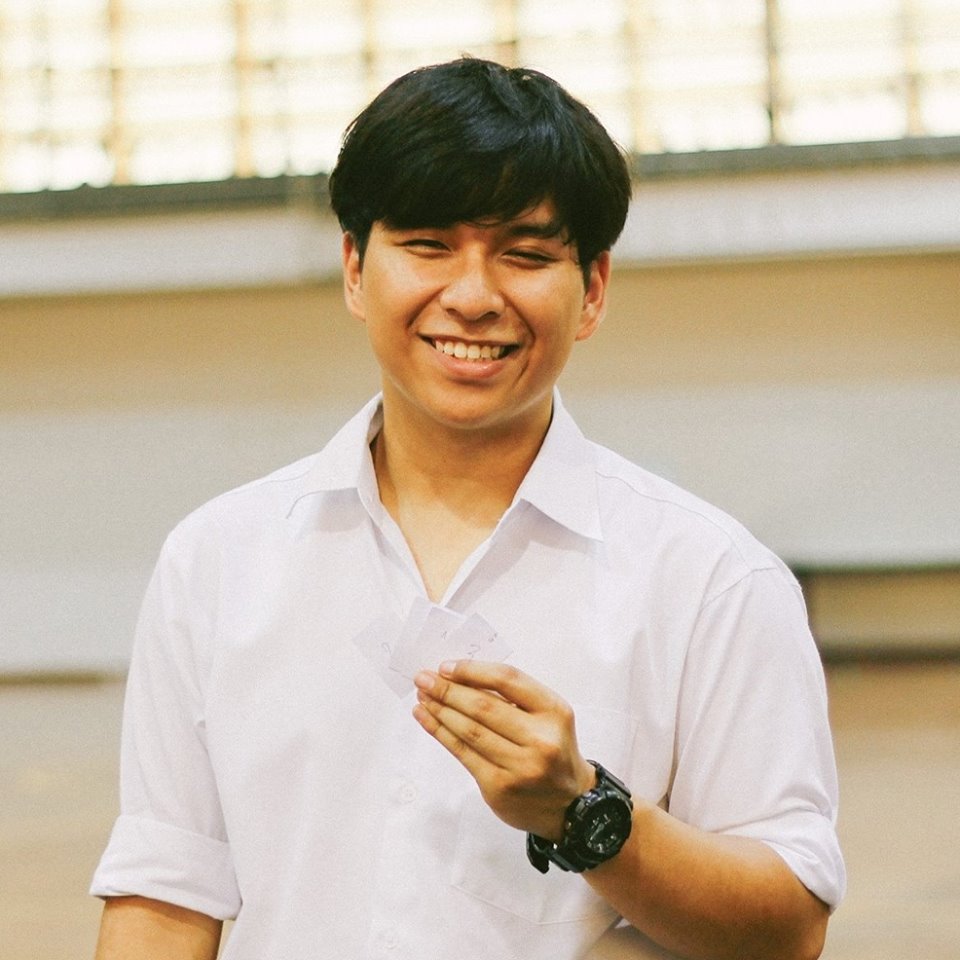}}]
{Karis Matchaparn} is currently studying in Undergraduate Program of Computer Engineering, King Mongkut's University of Technology Thonburi, Thailand. He is now working toward Deep Learning approach for EEG applications with BRAIN lab at School of Information Science and Technology (IST), Vidyasirimedhi Institute of Science and Technology (VISTEC), Thailand.
\end{IEEEbiography}
\vskip -2pt plus -1fil
\begin{IEEEbiography}[{\includegraphics[width=1.1in,height=1.7in,clip,keepaspectratio]{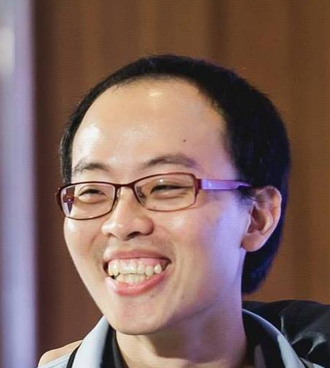}}]
{Tanaboon Tongbuasirilai} is a Ph.D. candidate in Media and Information Technology at Link\"{o}ping University, Sweden. He received B.Sc. in Mathematics from Mahidol University, Thailand.  In 2013, he recieved M.Sc. in Advanced Computer Graphics from Link\"{o}ping University, Sweden. His research interests lie in the intersection of computer graphics and image processing.
\end{IEEEbiography}
\vskip -2pt plus -1fil
\begin{IEEEbiography}[{\includegraphics[width=1in,height=1.25in,clip,keepaspectratio]{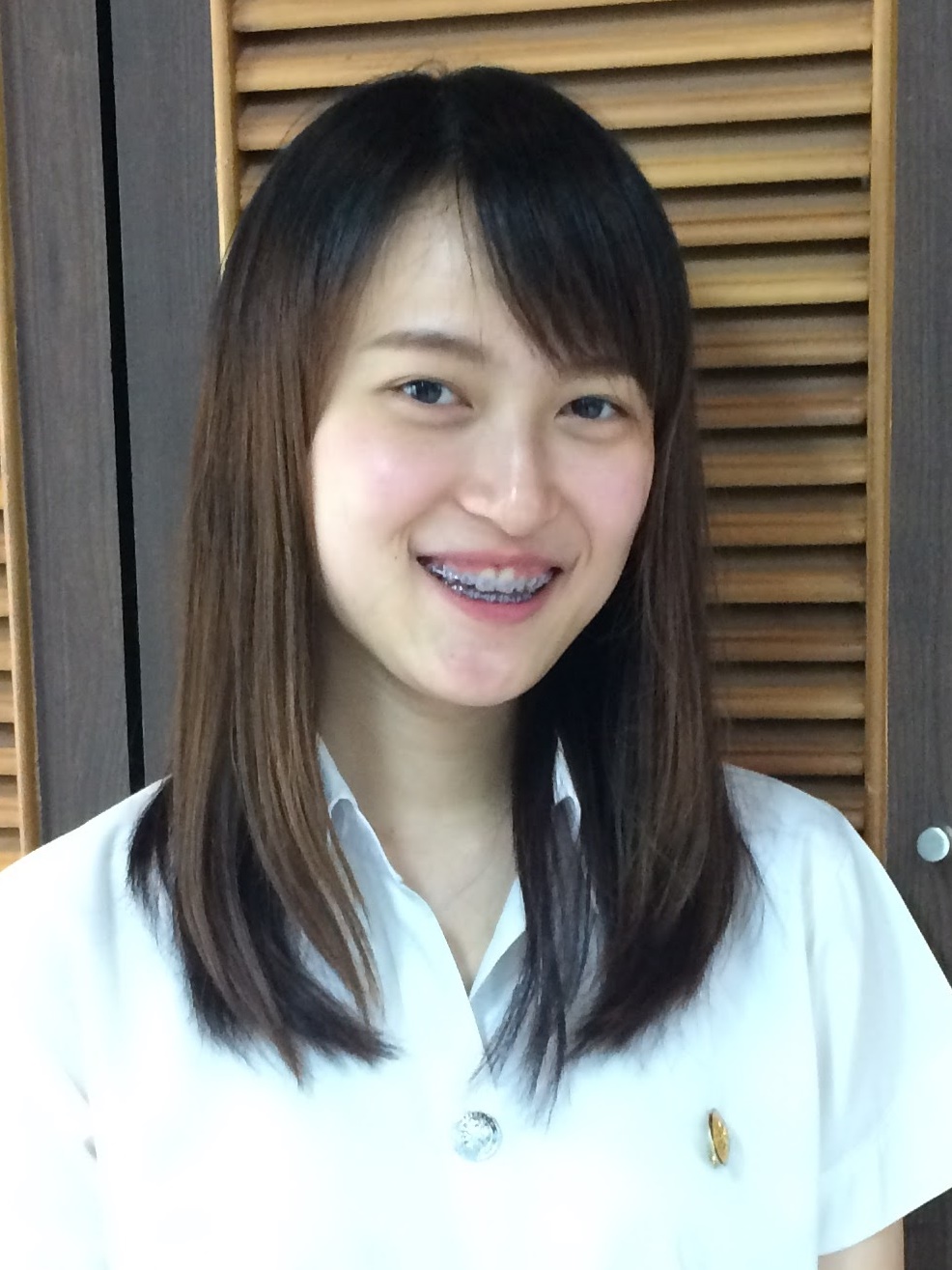}}]
{Nannapas Banluesombatkul} recieved the B.Sc. degree in Computer Science from Thammasat University, Thailand in 2017. She is currently a Research Assistant with Bio-inspired Robotics and Neural engineering (BRAIN) lab, School of Information Science and Technology at Vidyasirimedhi Institute of Science and Technology (VISTEC), Thailand. Her current research interests include biomedical signal processing and clinical diagnosis support system.
\end{IEEEbiography}
\vskip -2pt plus -1fil
\begin{IEEEbiography}[{\includegraphics[width=1.1in,height=1.6in,clip,keepaspectratio]{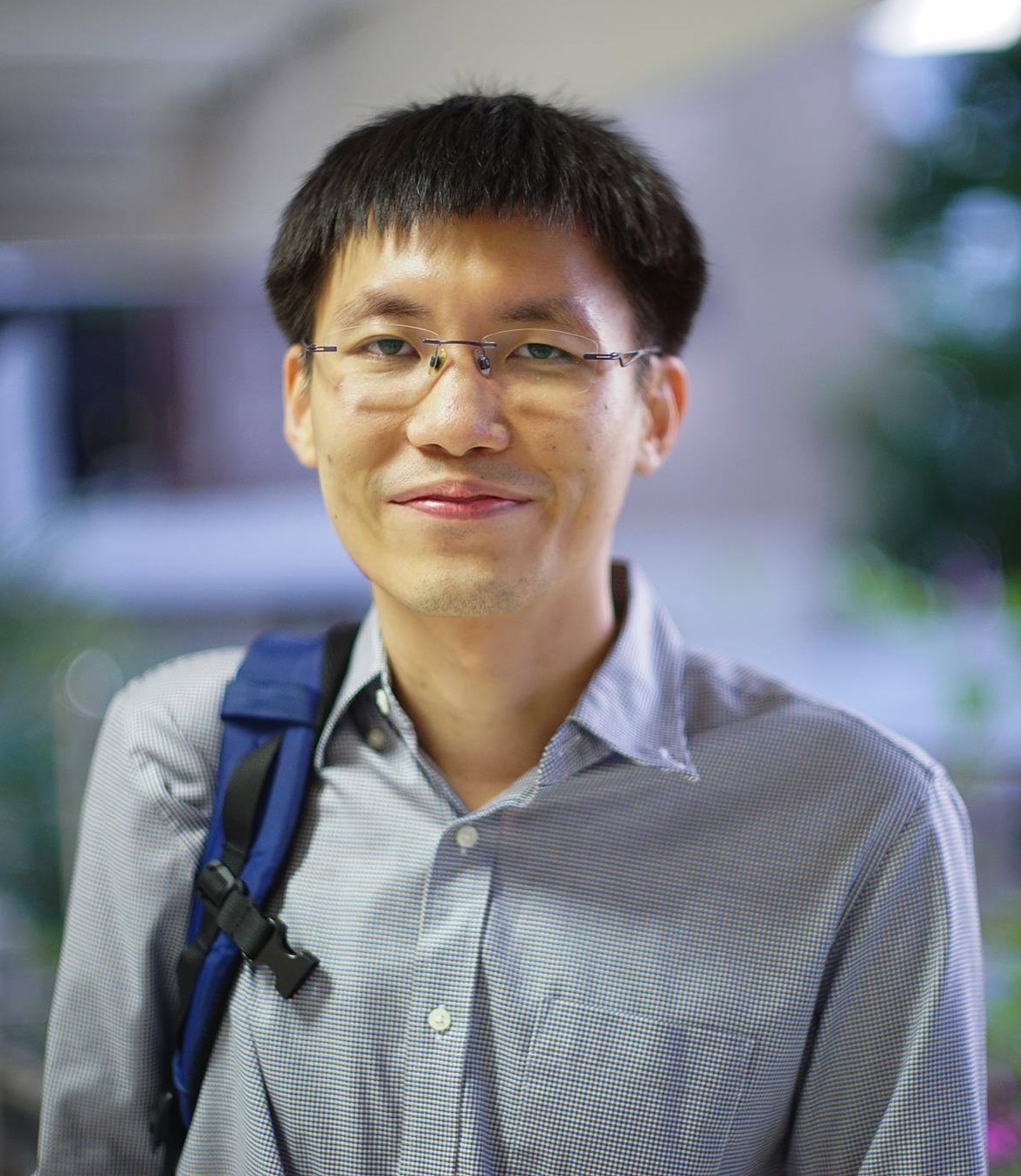}}]
{Ekapol Chuangsuwanich} received the B.S. and S.M. degree in Electrical and Computer Engineering from Carnegie Mellon University in 2008 and 2009. He then joined the Spoken Language Systems Group at MIT Computer Science and Artificial Intelligence Laboratory. He received his Ph.D. degree in 2016 from MIT. He is currently a Faculty Member of the Department of Computer Engineering at Chulalongkorn University. His research interests include machine learning approaches applied to speech processing, assistive technology, and health applications.
\end{IEEEbiography}
\end{document}